\documentclass[10pt,twocolumn,reqno,amsmath,amssymb,aps,prb]{revtex4-2}
\usepackage{graphicx}
\usepackage[dvipsnames]{xcolor}
\usepackage{hyperref}    
\usepackage{xcolor}
\usepackage{leftindex}
\usepackage{float}

\usepackage{dcolumn}% Align table columns on decimal point
\usepackage{bm}% bold math

\usepackage{dcolumn}% Align table columns on decimal point
\usepackage{bm}% bold math
\begin{document}

\preprint{APS/123-QED}

\title{Spin-caloritronic signatures of soft magnons in bilayer CrSBr}%

\author{Rob den Teuling$^1$}
\altaffiliation[Corresponding author:]{ r.denteuling@tudelft.nl.}

\author{Ping Tang$^{2}$}

\author{Gerrit E. W. Bauer$^{2}$}

\author{Yaroslav M. Blanter$^1$ \vspace{3mm}}

\affiliation{${}^1$ \hspace{-3mm} Kavli Institute of Nanoscience, Delft University of Technology, Lorentzweg 1, 2628 CJ, Delft, The Netherlands, \vspace{2mm}}%

\affiliation{${}^2$ \hspace{-3mm} WPI-AIMR \& IMR  \& CSIS, Tohoku University, 2-1-1 Katahira, Sendai 980-8577, Japan. 
\vspace{2mm}}

\date{\today}% It is always \today, today,

\begin{abstract}
Spin transport in magnetic insulators is often treated by assuming that magnons carry a fixed spin angular momentum of \(\hbar\), which does not hold in general, however. Here we calculate the magnon spin angular momentum of a layered antiferromagnet as a function of applied magnetic field and wave vector. We show that the triaxial anisotropy and intralayer dipolar interactions in bilayer CrSBr renormalize the magnon spin angular momentum, which diverges upon field-induced magnon softening. This divergence gives rise to a pronounced peak in the thermal spin Seebeck response and provides a clear spin-caloritronic signature of soft magnons.
\end{abstract}

\maketitle

%\tableofcontents

\section{\label{sec:level1}Introduction}

The spin current generated by thermal gradients in magnetic insulators is a central theme in  spin caloritronics. When injected into a metal contact, a transverse voltage builds up due to the inverse spin Hall effect \cite{Saitoh2006, Uchida2008, Bauer2012, Boona2014}, observed in a wide range of mainly oxidic magnetic insulators \cite{Uchida2010, Rezende2016, Wu2018, Holanda2017}. 

Single or multilayered two-dimensional (2D) van der Waals (vdW) magnets offer an attractive platform for spintronics. Spin transport as a function of external magnetic and electric fields and stacking geometry \cite{Gong2017, Gibertini2019, Huang2017} may provide valuable information on the carriers of spin and energy, \textit{viz}. the quanta of the spin waves or magnons. The relatively weak interlayer coupling in vdW multilayers causes magnetic properties to differ strongly from those of to bulk materials \cite{MacNeill2019, Bonilla2018}. The gate control of the interlayer exchange found in CrI\(_3\) \cite{huang2018} may be useful in energy-saving spintronic devices.

CrSBr is a particularly interesting vdW magnetic insulator with a tetragonal
crystal lattice structure. It displays intralayer ferromagnetic (FM) and interlayer antiferromagnetic (AFM) coupling \cite{Gong2019}, and strong triaxial magnetocrystalline anisotropy \cite{Yang2021, Ziebel2024, Krelle2025}. The bilayer has been accurately described in Refs. \cite{Teuling2,Rezende2026}. The presence of (anti)ferromagnetic resonance frequencies in the GHz range \cite{Rezende2026, Teuling2}, together with its air-stable nature \cite{Ziebel2024}, makes it a promising candidate for spintronic applications. Despite a growing body of work on magnetic order and excitations, the implications of the magnon mode structure for spin transport in CrSBr remain largely unexplored.

On a microscopic level, the spin-caloritronic response depends on the magnon band structure and the mode-dependent spin angular momentum (SAM) \cite{Bauer2023}. The magnon SAM is often assumed to be \(\hbar\) \cite{Bender2012, Rezende20162, Cornelissen2016} but this need not be the case in the presence of magnetic anisotropy and dipole-dipole interactions \cite{Kamra2017, Okuma2017, Bauer2023}, both of which play an important role in CrSBr \cite{Teuling2}.  Here, we investigate thermally induced and diffusive magnon spin transport governed by the spin Seebeck coefficient (SSC) in mono- and bilayer CrSBr, including the non-universal SAM.

The paper is organized as follows. In Sec. \hyperref[II]{II} we start from the Hamiltonians introduced in Ref. \cite{Teuling2} for mono- and bilayer CrSBr.  In Sec. \hyperref[III]{III} we calculate the magnon SAM as the expectation value of the spin operator for magnon eigenstates. In Sec. \hyperref[IV]{IV} we compute the SSC under in-plane external fields and gradients of temperature and chemical potential. Section \hyperref[V]{V} contains a summary and discussion.

\section{\label{II}Theoretical Framework}

Throughout this work, we adopt the model and material parameters from Ref. \cite{Teuling2}, where we already computed the magnon spectrum. To compute the magnon spin, we require here also the magnon wave functions. Figure \hyperref[fig5]{5} shows the top view of a CrSBr monolayer. The Hamiltonian for the monolayer (bilayer) is shown in Appendix \hyperref[appB]{B} (\hyperref[appC]{C}). As in our previous work \cite{Teuling2}, we model the monolayers as in-plane ferromagnets with a single spin per unit cell. Here we focus on the magnetodipolar terms. For a thin film with in-plane magnetization oriented along the $\hat{e}^{(i)}_\gamma$ axis, where \(\gamma\) is a Cartesian coordinate and \(i\) (\(=A,B\)) the layer index (see Appendix \hyperref[appB3]{B3} and \hyperref[appC3]{C3}) the dynamic dipolar field at site \(j\) is given by \cite{Teuling}
\begin{equation} \label{1}
\begin{aligned}
    \vec{B}_{dip,j,i} (\vec{k})= & - \mu_0 M_s f(k) m_{\beta,j,i} \hat{e}_\beta^{(i)} \\
    & - \mu_0 M_s \frac{k_{\alpha,i}^2}{k^2}(1-f(k))m_{\alpha,j,i} \hat{e}_\alpha^{(i)}, \\
\end{aligned}
\end{equation}
\noindent
where \(\mu_0\) is the vacuum permeability, \(M_s\) is the saturation magnetization, the \(\alpha\)(\(\beta\))-direction is the in-plane (out-of-plane) Cartesian component perpendicular to \(\gamma\), and  \(f(k) = (1-e^{-kd})/(kd)\), where \( k = |\vec{k}| \) and the thickness of the monolayer \(d \approx 
b/2\) \cite{Teuling2} with \(b\) the easy axis lattice vector. The magnetic moment \( \vec{m}_j = [m_{\alpha,j},m_{\beta,j},m_{\gamma,j}]^T = \gamma \hbar \vec{S}_j /M_s \), where \(\gamma\) is the gyromagnetic ratio (not to be confused with subscript \(\gamma\) for the Cartesian coordinate). The Zeeman interaction of the magnetic moments with the dipolar field can be written as a spin Hamiltonian (see Eqs. (\hyperref[B1]{B1}) and (\hyperref[C1]{C1})), 
\begin{equation} \label{2}
\begin{aligned}
    H_{dip,i} (\vec{k}) & = \sum_j \Bigg( \frac{1}{2}\mu_0 \frac{\gamma \hbar M_s}{S} f(k) (S_{j,i}^{(\beta)})^2 \\
    & + \frac{1}{2}\mu_0 \frac{\gamma \hbar M_s}{S} \frac{k_{\alpha,i}^2}{k^2}(1-f(k)) \left(S_{j,i}^{(\alpha)}\right)^2 \Bigg). \\
\end{aligned}
\end{equation}

\subsection{Monolayer}

The monolayer magnon Hamiltonion can be derived from the spin Hamiltonian in Appendix \hyperref[appB1]{B1} via the HP transformation \cite{HolsteinPrimakoff1940}
\begin{equation} \label{3}
\begin{aligned}
S_j^{(\gamma)} &= S - \hat{a}_j^\dagger \hat{a}_j, \\
S_j^{(\alpha)} &= \sqrt{\frac{S}{2}}(\hat{a}_j + \hat{a}_j^\dagger), \\
S_j^{(\beta)} &= -i \sqrt{\frac{S}{2}}(\hat{a}_j - \hat{a}_j^\dagger),
\end{aligned}
\end{equation}

\noindent
where  \(\alpha\) and \(\beta\) are the transverse components to \(\gamma\). To lowest order in the magnon number with \(\hat{a}_i = \frac{1}{\sqrt{N}} \sum_{{k}} \hat{a}_{k} e^{i \vec{k} \cdot \vec{r}_i}\) the Hamiltonian reads
\begin{equation} \label{4}
H^{(l)} = - \sum_{k} \Big[A^{(l)}_k \hat{a}_k^{\dagger} \hat{a}_k + \frac{B^{(l)}_k}{2}\, \left( \hat{a}_k \hat{a}_{-k} + \hat{a}_k^{\dagger} \hat{a}_{-k}^\dagger \right) + C^{(l)}_k 
\Big],
\end{equation}

\noindent
where \(l\) indicates the magnetic (parallel, antiparallel, or canted) configuration and dependence on the direction and magnitude of the external magnetic field (see Sec. \hyperref[III]{III}). The functions \(A_k\) and \(B_k\) are shown in Appendix \hyperref[appB4]{B4} and \(\sum_k C^{(l)}_k\) contributes a constant shift that can be disregarded. We introduce the Bogoliubov transformation for the monolayer \cite{Bogoliubov1947} \(\Phi_1 = T_1 \Psi_1\), where
\begin{equation} \label{5}
\Psi_1 =
\begin{pmatrix}
\hat{a}_k \\ \hat{a}_{-k}^\dagger \\
\end{pmatrix}, \quad
\Phi_1 =
\begin{pmatrix}
\hat{\alpha}_k \\ \hat{\alpha}_{-k}^\dagger
\end{pmatrix}, \quad 
T_1 = 
\begin{bmatrix}
u_k & v_k \\
v_k & u_k \\
\end{bmatrix}.
\end{equation}

\noindent
 Eq. (\hyperref[4]{4}) reads
 \begin{equation} \label{6}
H^{(l)} = \frac{1}{2} \Psi_1^\dagger \mathcal{H}^{(l)} \Psi_1 + \text{constant},
\end{equation}
where 
\begin{equation} \label{7}
\mathcal{H}^{(l)} = 
\begin{bmatrix}
-A^{(l)}_k & -B^{(l)}_k \\
-B^{(l)}_k & -A^{(l)}_k \\
\end{bmatrix}
\end{equation}
and the factor \(\frac{1}{2}\) corrects for double counting. 
The elements of both the original (\(\hat{a}\)) and quasiparticle (\(\hat{\alpha}\)) operator vectors obey the bosonic commutation relations \([\Psi_{1,i}, \Psi_{1,j}^\dagger] = 
[\Phi_{1,i}, \Phi_{ ,j}^\dagger] =
\Sigma_{1,ij} \) \cite{Okuma2018}, where
\begin{equation} \label{8}
\Sigma_1 = 
\begin{bmatrix}
1 & 0 \\
0 & -1 \\
\end{bmatrix}.
\end{equation}

\noindent
The eigenvalues of \(\Sigma_1 \mathcal{H}^{(l)}\) give the magnon spectrum \(\omega^{(l)}_\alpha = \sqrt{\left(A^{(l)}_k\right)^2 - \left(B^{(l)}_k\right)^2}\) \cite{Colpa1978}, while 
\begin{equation} \label{9}
    T_1\Sigma_1 \mathcal{H}^{(l)} T_1^{-1} = \Sigma_1 \mathcal{E}^{(l)}.
\end{equation} 
with \(\mathcal{E}^{(l)}=\text{diag}(\omega^{(l)}_\alpha, \omega^{(l)}_\alpha)\). Moreover, the $j$th column of $T_1^{-1}$ is an eigenvector of $\Sigma_1 \mathcal{H}^{(l)}$ so that
\begin{equation} \label{10}
    \Sigma_1 \mathcal{H}^{(l)}
    \begin{pmatrix}
        u_k \\
        -v_k
    \end{pmatrix}
    =
    \omega_\alpha^{(l)}
    \begin{pmatrix}
        u_k \\
        -v_k
    \end{pmatrix}.
\end{equation}

\subsection{Bilayer}

In the bilayer each spin is transformed in its respective static magnetization frame via
\vspace{-2mm}
\begin{equation} \label{11}
\begin{minipage}{0.45\linewidth}
\[
\begin{aligned}
S_j^{(\gamma,A)} &= S - \hat{a}_j^\dagger \hat{a}_j, \\
S_j^{(\alpha,A)} &= \sqrt{\frac{S}{2}}(\hat{a}_j + \hat{a}_j^\dagger), \\
S_j^{(\beta,A)} &= -i \sqrt{\frac{S}{2}}(\hat{a}_j - \hat{a}_j^\dagger),
\end{aligned}
\]
\end{minipage}
\hspace{1em}
\begin{minipage}{0.45\linewidth}
\[
\begin{aligned}
S_j^{(\gamma,B)} &= S - \hat{b}_j^\dagger \hat{b}_j, \\
S_j^{(\alpha,B)} &= \sqrt{\frac{S}{2}}(\hat{b}_j + \hat{b}_j^\dagger), \\
S_j^{(\beta,B)} &= -i \sqrt{\frac{S}{2}}(\hat{b}_j - \hat{b}_j^\dagger),
\end{aligned}
\]
\end{minipage}
\end{equation}

\noindent
where \(\hat{a}_j^{(\dagger)}\) and \(\hat{b}_j^{(\dagger)}\) are the magnon operators for layers \(A\) and \(B\), respectively. The Bogoliubov transformation for the bilayer is \(\Phi_2 = T_2 \Psi_2\), where
\begin{equation} \label{12}
\Psi_2 =
\begin{pmatrix}
\hat{a}_k \\
\hat{b}_k \\
\hat{a}_{-k}^\dagger \\
b_{-k}^\dagger
\end{pmatrix}\hspace{-1mm},
\quad
\hspace{-3mm}\Phi_2 =
\begin{pmatrix}
\hat{\alpha}_k \\
\hat{\beta}_k \\
\hat{\alpha}_{-k}^\dagger \\
\hat{\beta}_{-k}^\dagger
\end{pmatrix}\hspace{-1mm},
\quad
\hspace{-3mm}T_2 = 
\begin{bmatrix}
u_a^{(\alpha)} & \hspace{-1mm}u_b^{(\alpha)} & \hspace{-1mm}v_a^{(\alpha)} & \hspace{-1mm}v_b^{(\alpha)} \\
u_a^{(\beta)} & \hspace{-1mm}u_b^{(\beta)} & \hspace{-1mm}v_a^{(\beta)} & \hspace{-1mm}v_b^{(\beta)} \\
v_a^{(\alpha)} & \hspace{-1mm}v_b^{(\alpha)} & \hspace{-1mm}u_a^{(\alpha)} & \hspace{-1mm}u_b^{(\alpha)} \\
v_a^{(\beta)} & \hspace{-1mm}v_b^{(\beta)} & \hspace{-1mm}u_a^{(\beta)} & \hspace{-1mm}u_b^{(\beta)}
\end{bmatrix}\hspace{-1mm},
\end{equation}
with \([\Psi_{2,i}, \Psi_{2,j}^\dagger] = 
[\Phi_{2,i}, \Phi_{2,j}^\dagger] =
\Sigma_{2,ij} \) and
\begin{equation} \label{13}
\Sigma_2 = 
\begin{bmatrix}
1 & 0 & 0 & 0 \\
0 & 1 & 0 & 0 \\
0 & 0 & -1 & 0 \\
0 & 0 & 0 & -1 \\
\end{bmatrix}.
\end{equation} 
This leads to an equation analogous to (\ref{9}) with eigenvalues in  \(\mathcal{E}^{(l)} = \text{diag}(\omega_\alpha^{(l)},\omega_\beta^{(l)},\omega_\alpha^{(l)},\omega_\beta^{(l)})\) that solve an equation similar to (\ref{10}):
\begin{equation} \label{14}
    \Sigma_2 \mathcal{H}^{(l)}\begin{pmatrix}
        u_a^{m} \\
        u_b^{m} \\
        - v_a^{m} \\
        - v_b^{m} \\
\end{pmatrix} =\omega^{(l)}_m \begin{pmatrix}
        u_a^{m} \\
        u_b^{m} \\
        - v_a^{m} \\
        - v_b^{m} \\
\end{pmatrix},
\end{equation}

\noindent
where \(m = \alpha,\beta\) corresponding to the two magnon bands of the bilayer. We show \(\mathcal{H}^{(l)}\) for all phases in the bilayer in Appendix \hyperref[appC4]{C4}.

\section{\label{III}Spin Angular Momentum}

In this section, we evaluate the non-universal magnon SAM in both the mono- and bilayer by computing the expectation value of the static magnetization in the diagonal Bogoliubov basis.

\subsection{Monolayer}

The SAM in the monolayer of a single magnon branch \(\alpha\) with its spins quantized along the equilibrium magnetization (along the \(\gamma\)-axis)  \cite{Okuma2017}
\begin{equation} \label{15}
    \Delta S^{(\gamma)}_{\alpha} = \langle 1_\alpha|S^{(\gamma)}|1_\alpha \rangle - \langle0|S^{(\gamma)}|0\rangle,
\end{equation}
where \(|0\rangle\) is the vacuum with \(\hat{\alpha}_k|0\rangle = 0\) and \(|1_\alpha \rangle = \alpha^\dagger_k|0 \rangle\). Combining Eqs. (\hyperref[3]{3}), (\hyperref[5]{5}), and Eq. (\hyperref[15]{15}) leads to \(\Delta S^{(\gamma)}_{\alpha} = -|u|^2\) in units of \(\hbar\). Combining Eq. (\hyperref[10]{10}) together with the bosonic commutation condition \(|u|^2 - |v|^2 = 1\) (\(\hat{e}_j^T \Sigma \hat{e}_j=1\))  we find
\begin{equation} \label{16}
    \Delta S^{(\gamma,l)}_\alpha = |u^{(l)}|^2 = \frac{1}{1-\left(\frac{\omega^{(l)}_\alpha + A^{(l)}_k}{B^{(l)}_k}\right)^2}.
\end{equation}
In the canted configuration (see Appendix \hyperref[appB3]{B3}) the SAM along the \(x\)- and \(z\)-axes are \(- |u^{(l)}|^2 \sin\theta\) and \(- |u^{(l)}|^2 \cos\theta\), respectively.

An external field along the easy axis (\(b\)) of the monolayer leads to the collinear configuration (\(\gamma=z, l = \text{f}\)) in which the magnetization direction is along \(z\). A field along the intermediate axis (\(a\)) with \(B_0 < B_{\text{sat}}\) cants the spins at an angle between  the \(b\)- and \(a\)-axes (\(l = \text{m,c}\)), while for \(B_0 \geq B_{\text{sat}}\) the magnetization is along the \(a\)-axis (\(\gamma=x,l = \text{m,s}\)). Appendix \hyperref[appB4]{B4} shows the matrix elements of \(\mathcal{H}^{(l)}\) and Fig. \hyperlink{fig1}{1} the field dependence of the SAM for the Kittel mode at \(k=0\).
\begin{figure} [ht] \label{fig1}

%\subfloat{%
  \hspace{-3mm}\includegraphics[clip,width=1\columnwidth]{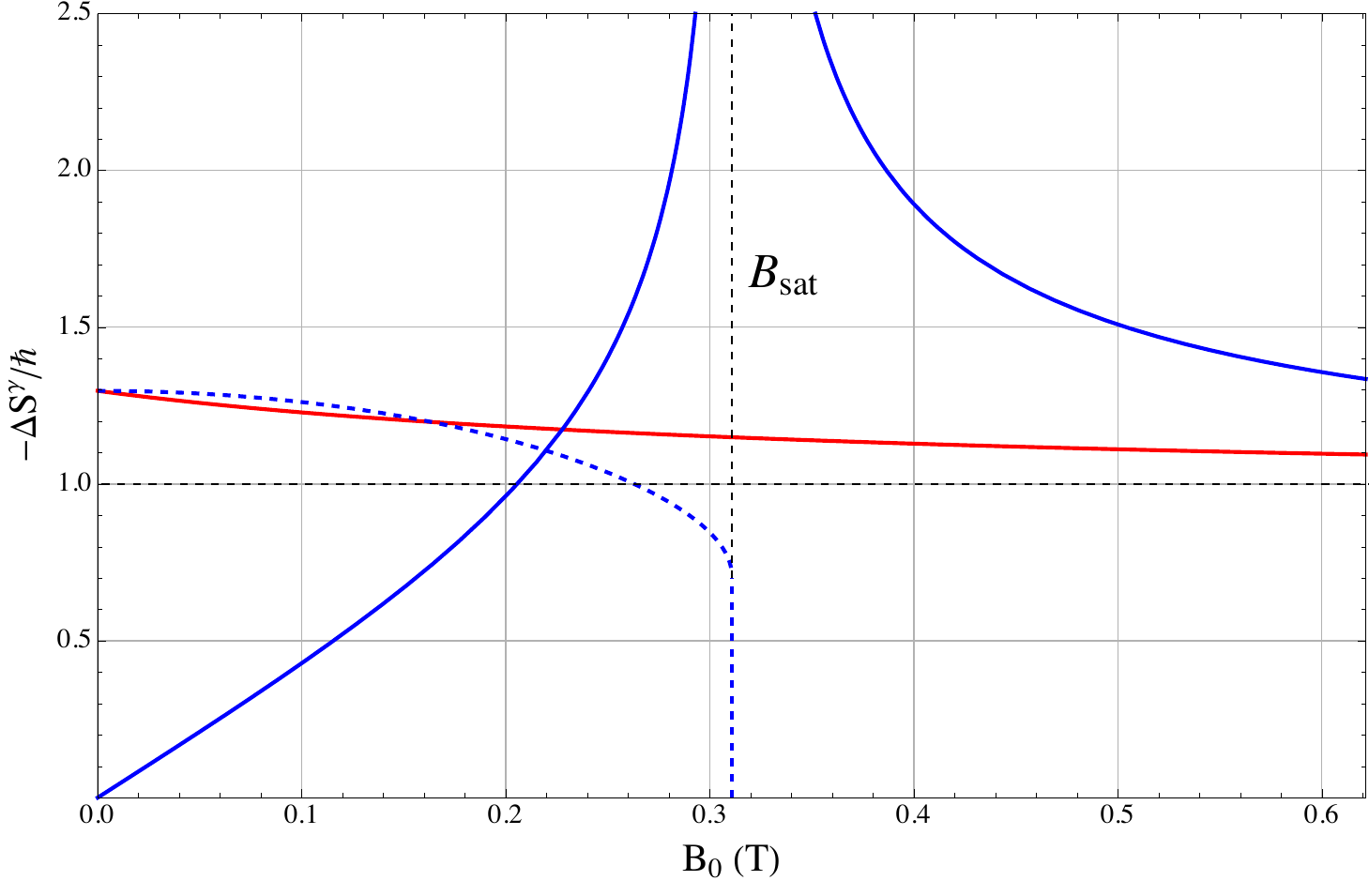}%
%}

\caption{ \small Calculated SAM of the Kittel mode \((k=0)\) in the monolayer as a function of the external magnetic field \(B_0\). The red curve shows the SAM along the easy axis ($\vec{B}_0 \parallel \hat{b}$) while the blue ones hold for $\vec{B}_0 \parallel \hat{a}$ in the canted (\(B_0 < B_{\text{sat}}\)) and saturated (\(B_0 \geq B_{\text{sat}}\)) phases: the solid (dotted) line shows the projection of the SAM along the $a$-axis ($b$-axis).}
\end{figure}

The red curve corresponds to collinear spin and external field along the easy axis (\(b\)). Here a SAM larger than \(\hbar\) indicates the degree of elliptical polarization \cite{Kamra2016} by the triaxial anisotropy and intralayer dipole interactions \cite{Teuling2}. In the limit of a large applied magnetic field field \(|\Delta S ^{(\gamma)}|\rightarrow \hbar\) .

The blue curves are the results for an external field along the intermediate axis (\(a\)). The solid blue curve is the projection of the SAM along \(a\), and the dotted curve along \(b\), perpendicular to the external field. For fields approaching the saturation value the SAM increases strongly and ultimately diverges when the magnon frequency softens to zero, reflecting an increasingly elliptical precession \cite{Bauer2023}. At the critical soft point (\(\omega \rightarrow 0\)) the magnon becomes linearly polarized. The amplitude diverges because anisotropy and magnetic field torques cancel at this point. However, close to \(B_{\mathrm{sat}}\) the small-amplitude assumption underlying the non-interacting magnon approximation and the validity of the Holstein-Primakoff expansion break down.  Far above \(B_{\mathrm{sat}}\), the SAM approaches \(\hbar\).

\subsection{Bilayer}

In the bilayer with parallel (\(+\)) or antiparallel (\(-\)) spin alignment, the SAM of mode \(m\) (\(=\alpha,\beta\))  spins read
\begin{equation} \label{17}
\begin{aligned}
    \Delta S^{(\gamma)}_{m} & = 
     \langle 1_m|S^{(\gamma,A)} \pm S^{(\gamma,B)}|1_m \rangle \\
     & - \langle0|S^{(\gamma,A)} \pm S^{(\gamma,B)}|0\rangle,
\end{aligned}
\end{equation}

\noindent
where in the vacuum \(\hat{m}_k|0\rangle = 0\) and \(|1_m \rangle = \hat{m}^\dagger_k|0 \rangle\), with \(\hat{m}_k^{(\dagger)}\) the quasi-particle operator from Eq. (\hyperref[12]{12}). From Eqs. (\hyperref[11]{11}), (\hyperref[12]{12}) and (\hyperref[17]{17}) it follows that \(\Delta S^{(\gamma,l)}_{{\alpha}} = -(|u^{(\alpha,l)}_a|^2 \pm |u^{(\alpha,l)}_b|^2)\) and \(\Delta S^{(\gamma,l)}_{\beta} = -(|u^{(\beta,l)}_a|^2 \pm |u^{(\beta,l)}_b|^2)\) in units of \(\hbar\), for which we combine Eq. (\hyperref[14]{14}) with the bosonic commutation relation \(|u|^2 - |v|^2 = 1\) (\(\hat{e}_j^T \Sigma \hat{e}_j=1\)). 

In the canted configuration (see Appendix \hyperref[appC]{C}) the \(z\)-components cancel by symmetry so that the effective SAM along the \(x\)-axis is \(-\sin\theta(|u^{(\alpha,l)}_a|^2 + |u^{(\alpha,l)}_b|^2)\) and \(-\sin\theta(|u^{(\beta,l)}_a|^2 + |u^{(\beta,l)}_b|^2)\) for the \(\alpha\) and \(\beta\) modes, respectively.

In the bilayer we distinguish four phases depending on the direction and magnitude of the applied field. When a field along the easy axis (\(b\)) stays below the spin-flip transition (\( B_0 < B_{crit}^{flip}\)), the AFM phase (\(\gamma = \pm z,l = \text{AFM}\)) is stable. Here, the spin are aligned anti-parallel along the easy axis. When \( B_0 \geq B_{crit}^{flip}\) the magnetization of both layers are parallel along the external field (\(\gamma = z,l = \text{FM}\)). Appendix \hyperref[appC4]{C4} derives \(\mathcal{H}^{(l)}\) in Eq. (\hyperref[14]{14}) for the AFM and FM phases in the bilayer, while Fig. \hyperlink{fig2a}{2(a)} shows the corresponding SAM as a function of the external magnetic field applied along the easy axis.
\begin{figure} [ht] \label{fig2}

%\subfloat[]{%
  \hspace{-3mm}\includegraphics[clip,width=1\columnwidth]{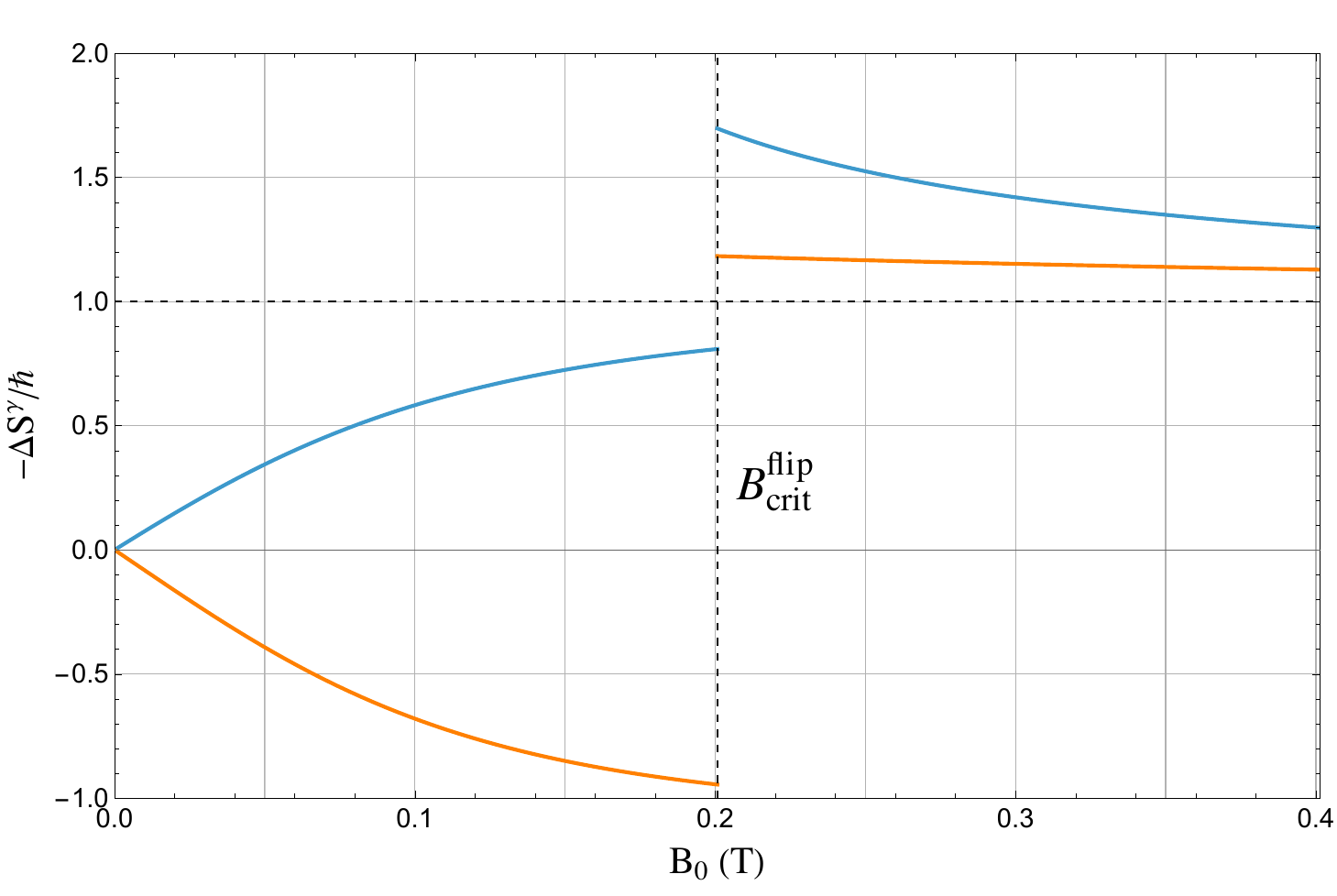}%
%}\label{fig2a}

%\subfloat[]{%
  \hspace{-3mm}\includegraphics[clip,width=0.97\columnwidth]{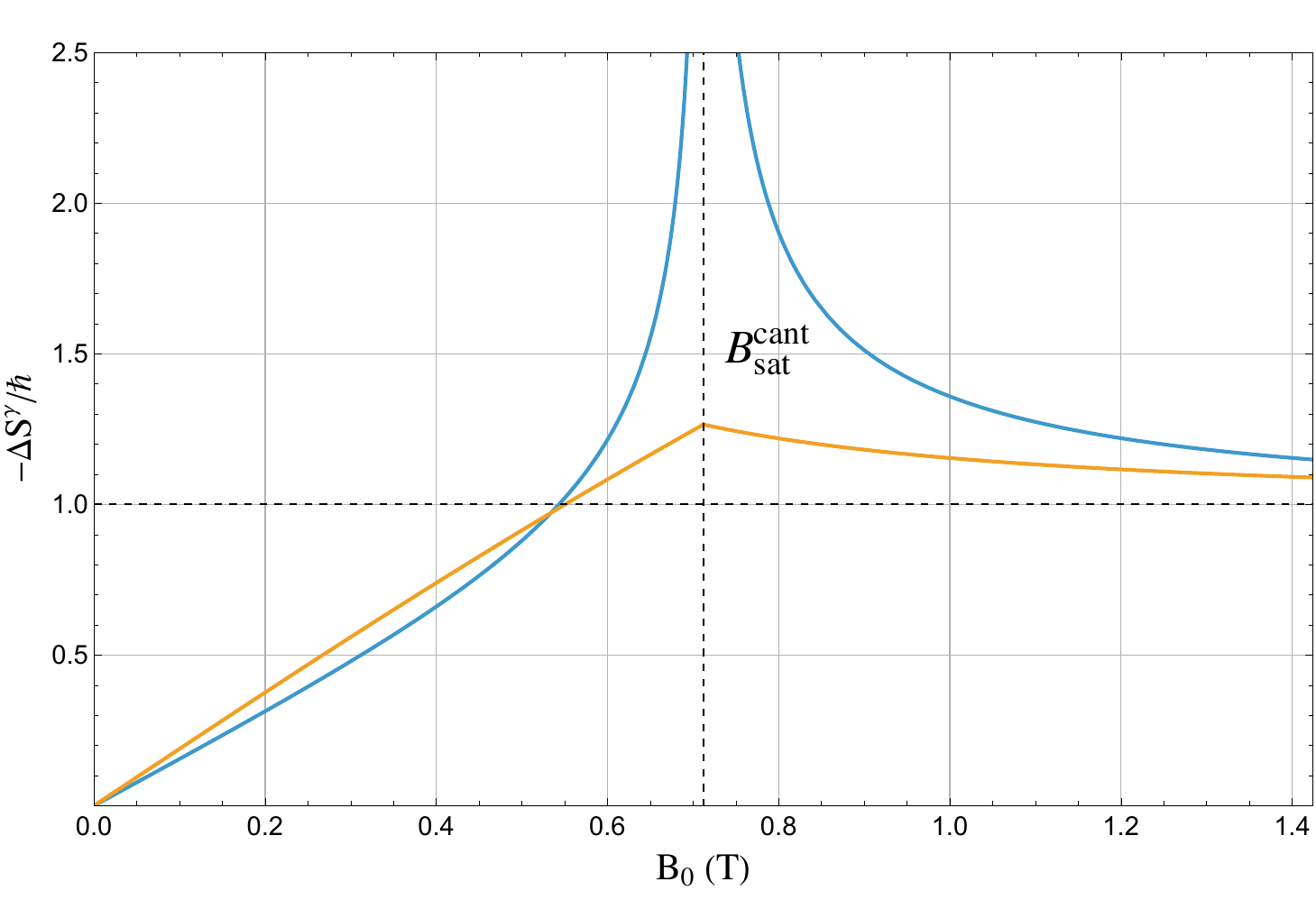}%
%}
\label{fig2b}

\caption{ \small SAM of the magnon modes at \(k=0\) in the bilayer as a function of the external magnetic field \(B_0\). The lower (higher) frequency band is indicated in blue (orange). (a) SAM along the easy axis (\(b\)) in the AFM (\(B_0 < B^{\text{flip}}_{\text{crit}}\)) and FM (\(B_0 \geq B^{\text{flip}}_{\text{crit}}\)) phases ($\vec{B}_0 \parallel \hat{b}$). (b) SAM along the intermediate axis (\(a\)) in the canted (\(B_0 < B^{\text{cant}}_{\text{sat}}\)) and saturated (\(B_0 \geq B^{\text{cant}}_{\text{sat}}\)) phases ($\vec{B}_0 \parallel \hat{a}$).}
\end{figure}
In the AFM regime, the two magnon branches carry opposite SAM \cite{Hoogeboom2021}. For small external fields the SAM is close to zero due to canceling \(z\)-components of comparable magnitude of the magnetization. At \(B_0 = B^{\text{flip}}_{\text{crit}}\), the system undergoes a spin-flip transition to the FM phase. In both phases the anisotropies cause deviations from \(\hbar\) as in the monolayer for \(\vec{B}_0 \parallel \hat{b}\).

Applying a field along the intermediate axis (\(a\)) symmetrically cants the spins from the \(\pm b\)-axes to the \(a\)-axis (\(B_0 < B^{\text{cant}}_{\text{sat}}, l = \text{b,c}\)), see Appendix  \hyperref[appC3]{C3}. At the saturation field \( B_0 \geq B^{\text{cant}}_{\text{sat}}\) all spins align with the \(a\)-axis (\(\gamma=x,l = \text{b,s}\)). Appendix \hyperref[appC4]{C4} derives \(\mathcal{H}^{(l)}\) in Eq. (\hyperref[14]{14}) for the canted and saturated phases in the bilayer, and Fig. \hyperlink{fig2b}{2(b)} shows the SAMs projected along the intermediate axis as a function of the external magnetic field. In the canted phase an increasing external field leads to the blue branch softening with enhanced ellipticity and SAM, analogous to the magnon mode in the monolayer. The algebraic divergence is again not physical but indicates the breakdown of the HP expansion. The second (orange curve)  branch remains at high-frequencies due to interlayer exchange with relatively small effects of the anisotropies for all field values, resulting in a weakly field-dependent SAM. At \(B_0 = B^{\text{cant}}_{\text{sat}}\), the system undergoes a transition to the saturated phase, where all spins become collinear with the external field. Beyond this point, the magnon spectrum hardens and the SAM of both modes gradually approaches \(\hbar\).

\section{\label{IV}Spin Seebeck Coefficients}

Based on the magnon bands (see Appendix \hyperlink{appB2}{B2} (\hyperlink{appC2}{C2}) for the monolayer (bilayer)) and SAMs in Figs. \hyperlink{fig1}{1}, \hyperlink{fig2a}{2(a)} and \hyperlink{fig2b}{2(b)}, we compute here the SSC by the linearized Boltzmann equation in the constant relaxation time approximation. 

Let \( n_k = n(\vec{r}, \vec{k},t)\) be the non-equilibrium distribution function of the magnon occupation numbers. The magnon spin current \cite{Rezende20162} in 2D reads
\begin{equation} \label{18}
    \vec{J}_S = \frac{1}{(2 \pi)^2}\int \Delta S_k \vec{v}_k(n_k - n^{(0)}_k)d^2k,
\end{equation}
where \(\Delta S_k\) the SAM from Sec. \hyperlink{III}{III},
$\vec{v}_k= d\omega_{\vec{k}}/d\vec{k}$
the magnon group velocity and \(n^{(0)} = (e^{\sigma}-1)^{-1}\) the occupation number at thermal equilibrium from the Bose-Einstein distribution with
\(\sigma = \hbar \omega_{\vec{k}} /(k_B T)\). 
Substituting the magnon distribution function 
\(n_k\) obtained from the linearized Boltzmann equation (\(n_k = n^{(0)} - \tau_k \Vec{v}_k\cdot\nabla_rn_k\)) for a constant relaxation time \(\tau\) \cite{Rezende20162}, the spin transport along \(\eta\) (\(=x,z\)) excited by a temperature gradient \(\partial T/\partial \eta\) reads
\begin{equation}\label{19}
\begin{aligned}
        J^{(\eta,\nabla T)}_{S} & = -\frac{\tau}{(2 \pi)^2}\int \Delta S_k  v^2_{k,\eta}\frac{\partial n^{(0)}}{\partial T}d^2k \frac{\partial T}{\partial \eta} \\
        & \equiv - S_S^{(\nabla T)}(B_0) \frac{\partial T}{\partial \eta}, \\
\end{aligned}
\end{equation}

\noindent
where \(S_S^{(\nabla T)}(B_0)\) is the thermal SSC due to thermal gradients. Likewise, Taylor expanding around \(\varepsilon_{\Vec{k}}\) (\(=\hbar \omega_{\vec{k}}\)) for a small chemical potential \(\mu\) in \(n_{\mu}^{(0)} = (e^{\sigma_{\mu}}-1)^{-1}\) with
\(\sigma_{\mu} = (\varepsilon_{\vec{k}}-\mu) /(k_B T)\), the spin transport excited by a chemical potential gradient \(\partial \mu/\partial \eta\)  reads
\begin{equation} \label{20}
\begin{aligned}
        J^{(\eta,\nabla \mu)}_{S} & = -\frac{\tau}{(2 \pi)^2}\int \Delta S_k  v^2_{k,\eta}\frac{\partial n^{(0)}_{\mu}}{\partial \mu} \Bigg|_{\mu=0} d^2k \frac{\partial \mu}{\partial \eta} \\
        & \equiv - S_S^{(\nabla \mu)}(B_0) \frac{\partial \mu}{\partial \eta}, \\
\end{aligned}
\end{equation}

\noindent
where \(S_S^{(\nabla \mu)}(B_0)\) is the diffusive SSC due to magnon accumulation. Since the magnon relaxation time in CrSBr is not known, we adopt a value \(\tau \approx 10 \hspace{1mm}\mu\)s, which is typical for yttrium iron garnet films \cite{Jamison2019,Gu2025}. 

\subsection{Monolayer CrSBr}

Fig. \hyperlink{fig3}{3} shows the thermal SSC in the monolayer at \(T=5\) K.
\begin{figure}[ht] \label{fig3}
  %\subfloat{%
    \hspace{-3mm}\includegraphics[clip,width=1\columnwidth]{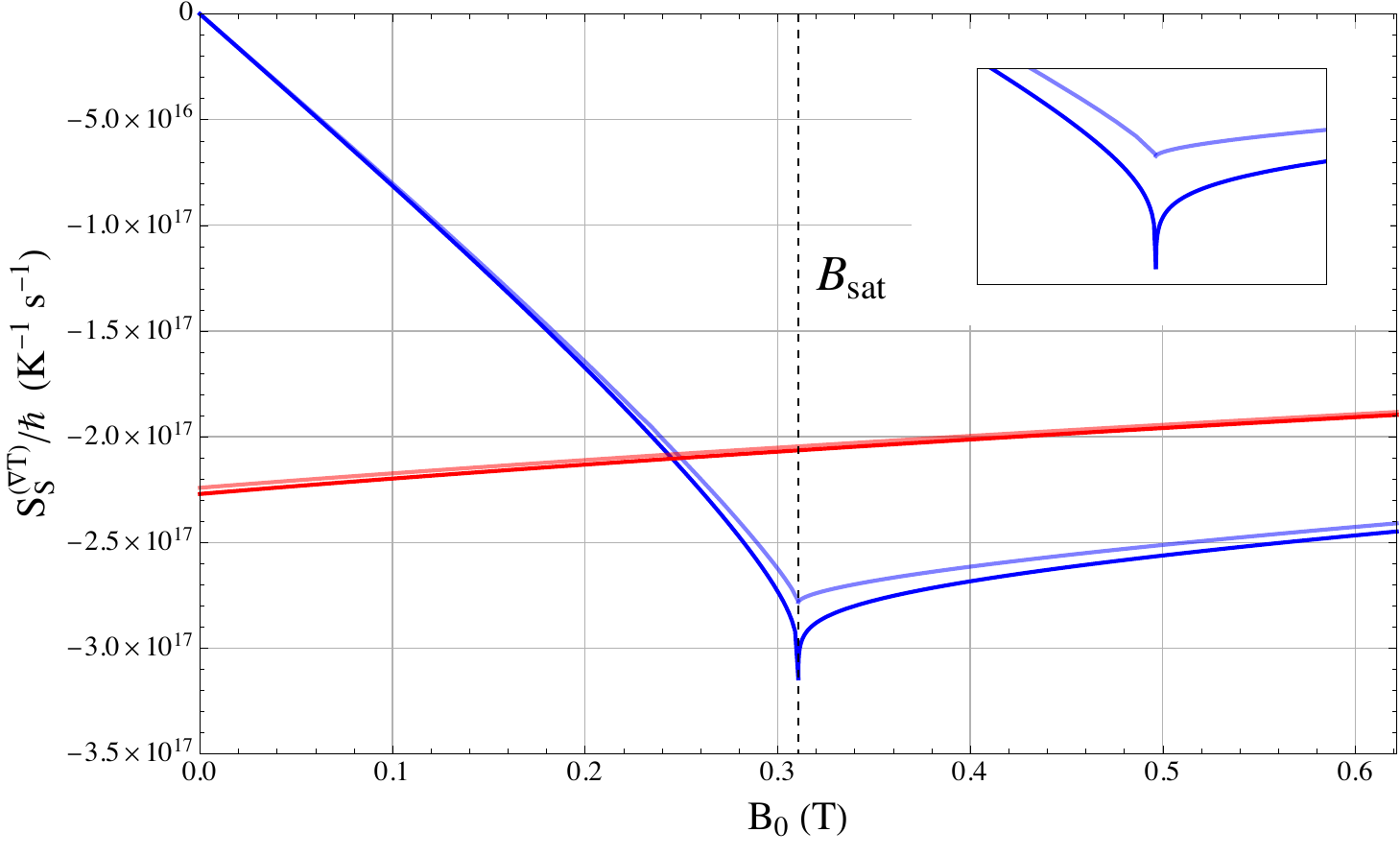}%
  %}
  \caption{\small Thermal SSC in a monolayer as a function of the external magnetic field \(B_0\). The transparent curves correspond to \(S_S^{(\nabla T)}\) for all SAM equal \(\hbar\). The red curves show \(S_S^{(\nabla T)}\) along the easy axis for $\vec{B}_0 \parallel \hat{b}$, the blue curves along the intermediate axis in the canted (\(B_0 < B_{\text{sat}}\)) and saturated (\(B_0 \geq B_{\text{sat}}\)) phases for $\vec{B}_0 \parallel \hat{a}$. The inset highlights a zoomed-in region around the peak of \(S_S^{(\nabla T)}\).}
\end{figure}
The red curve corresponds to magnetization and field parallel with the easy axis (\(b\)). The dependence on magnetic field is weak because the magnon SAM is nearly constant and close to $\hbar$.

The blue curve corresponds to the external field along the intermediate axis (\(a\)), which causes a logarithmic divergence of the thermal SSC around the saturation field $B_{\text{sat}}$ \cite{Bauer2023}. Imposing a constant \(\Delta S = -\hbar\) (transparent curves) suppresses the enhancement at \(B_{sat}\) where the curves meet linearly at a minimum and do not exhibit a peak. The predicted peak is therefore a spin-caloritronic signature of field-induced soft magnons. Fig. \hyperref[fig12]{12} in the SM illustrates the temperature dependence of the SSC at low temperatures, disregarding the temperature dependence of the magnetization.

Fig. \hyperlink{fig4}{4} shows the diffusive SSC in the monolayer at \(T=5\) K.
\begin{figure}[ht]
  %\subfloat{%
    \hspace{-3mm}\includegraphics[clip,width=1\columnwidth]{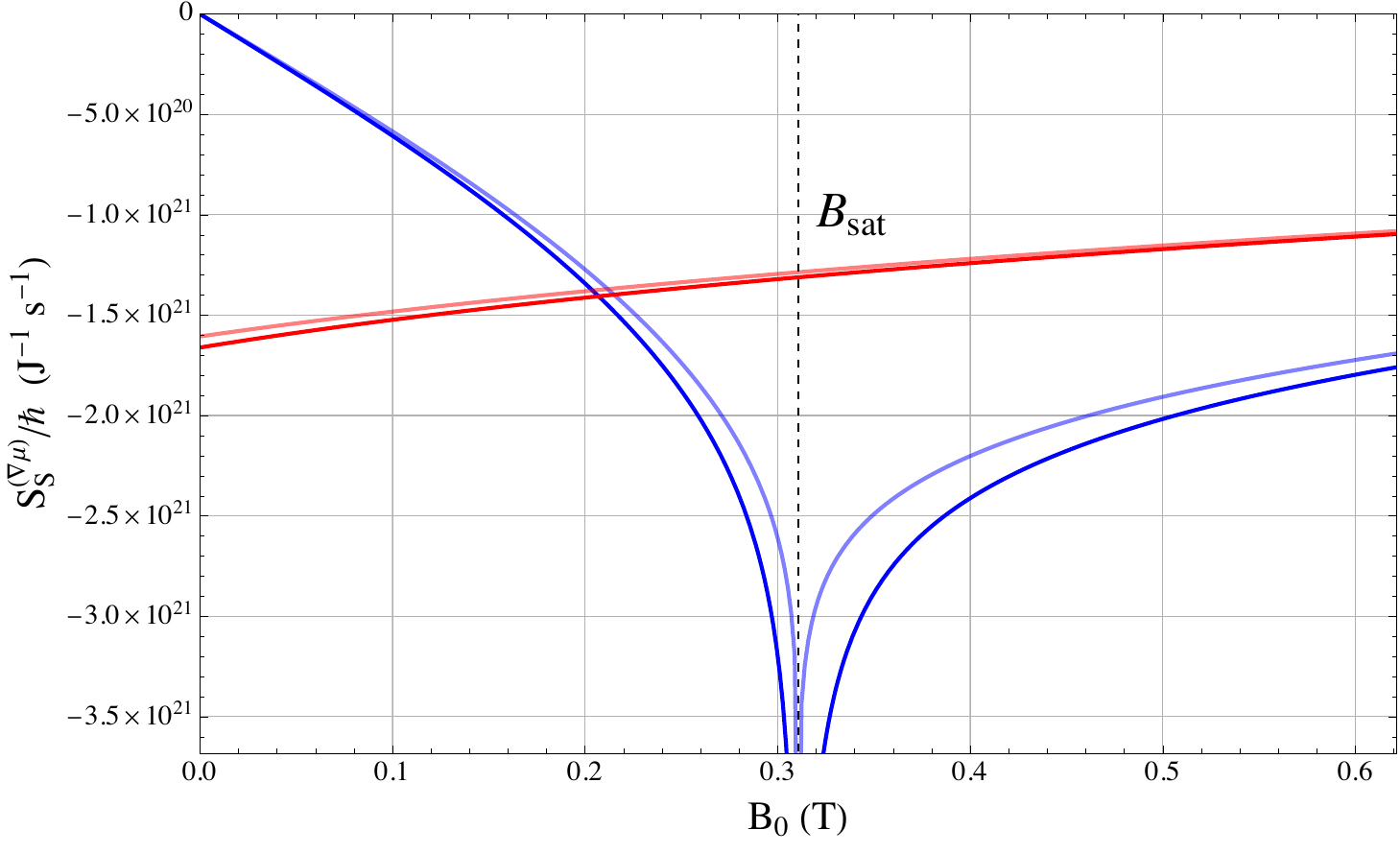}%
  %}
  \caption{\small Diffusive SSC in a monolayer as a function of the external magnetic field \(B_0\). The transparent curves correspond to \(S_S^{(\nabla \mu)}\) for all SAM equal \(\hbar\). The red curves show \(S_S^{(\nabla \mu)}\) along the easy axis for $\vec{B}_0 \parallel \hat{b}$, the blue curves along the intermediate axis in the canted (\(B_0 < B_{\text{sat}}\)) and saturated (\(B_0 \geq B_{\text{sat}}\)) phases for $\vec{B}_0 \parallel \hat{a}$.}
  \label{fig4}
\end{figure}
The red curves correspond to magnetization and field parallel with the easy axis (\(b\)). The dependence on the external field is weak for the same reason as in the thermal SSC.

The blue curves correspond to the external field along the intermediate axis (\(a\)), which now causes an enhancement of an already present peak of the diffusive SSC. Even for fixed magnon spin \(\Delta S= \hbar\), the diffusive SSC peaks near the softening field because the low-energy magnon occupation becomes highly sensitive to the chemical potential as the magnon gap closes. As \(\omega \rightarrow 0\), the low-energy magnon occupation grows large compared to \(S\). Consequently, higher-order magnon interactions neglected in linear spin wave theory become important and are expected to regularize the divergence.

Fig. \hyperref[fig13]{13} in the SM illustrates the temperature dependence at low temperatures. For completeness we show the magnon thermal conductivity for all phases in the monolayer in Fig. \hyperref[fig16]{16} in Appendix \hyperref[appE1]{E}.

\subsection{Bilayer CrSBr}

In this section we discuss the SSC in the bilayer. Fig. \hyperlink{fig5a}{5(a)} shows the thermal SSC of the bilayer in the AFM and FM phases for thermal gradient and external magnetic field applied along the easy axis (\(\vec{B}_0 \parallel \hat{b}\)) at \(T=5\) K. For fields below the spin-flip critical field $B^{\text{flip}}_{\text{crit}}$, the system is in the AFM phase. Both magnon branches contribute to the total thermal SSC proportional to their thermal occupation and partially cancel since they carry opposite SAM. Setting the SAM to \(\pm\hbar\) for the two modes appears to be a bad approximation at weak magnetic fields, but the error largely cancels in the total thermal SSC (black curves) apart from giving a small non-zero value. Above \(B_0 = B^{\text{flip}}_{\text{crit}}\) the sign of the SAM of both magnon branches is the same and close to $\hbar$ in both branches.  the field dependence of the thermal SSC remains relatively weak. The thermal SSC of the bilayer in the AFM and FM phases shows a discontinuity at the spin-flip field, but no divergence. 

Fig. \hyperlink{fig5b}{5(b)} shows the thermal SSC of the bilayer in the canted and saturated phases for thermal gradient and external magnetic field applied along the intermediate axis (\(\vec{B}_0 \parallel \hat{a}\)) at \(T=5\) K. When the field approaches \(B^{\text{cant}}_{\text{sat}}\), the frequency of lower mode approaches zero, leading to a divergence in the SAM. As in the monolayer, this strongly affects the thermal SSC since small-\(k\) magnons dominate spin transport. The thermal SSC of the high-frequency mode displays a pronounced kink at the phase transition.  In the saturated phase ($B_0 \geq B^{\text{cant}}_{\text{sat}}$), the spins are forced to become parallel and the soft magnon mode hardens again. At high fields the SAM of both modes approach \(\hbar\). As with the canted phase in the monolayer, setting the SAM to \(\hbar\) (semi-transparent curves) underestimates the thermal SSC and does not produce the peak at \(B_{\text{sat}}^{\text{cant}}\). Fig. \hyperref[fig14]{14} shows the total thermal SSC for various temperatures in all phases in the bilayer.

\begin{figure} [ht] \label{fig5}

%\subfloat[]{%
  \hspace{-3mm}\includegraphics[clip,width=1\columnwidth]{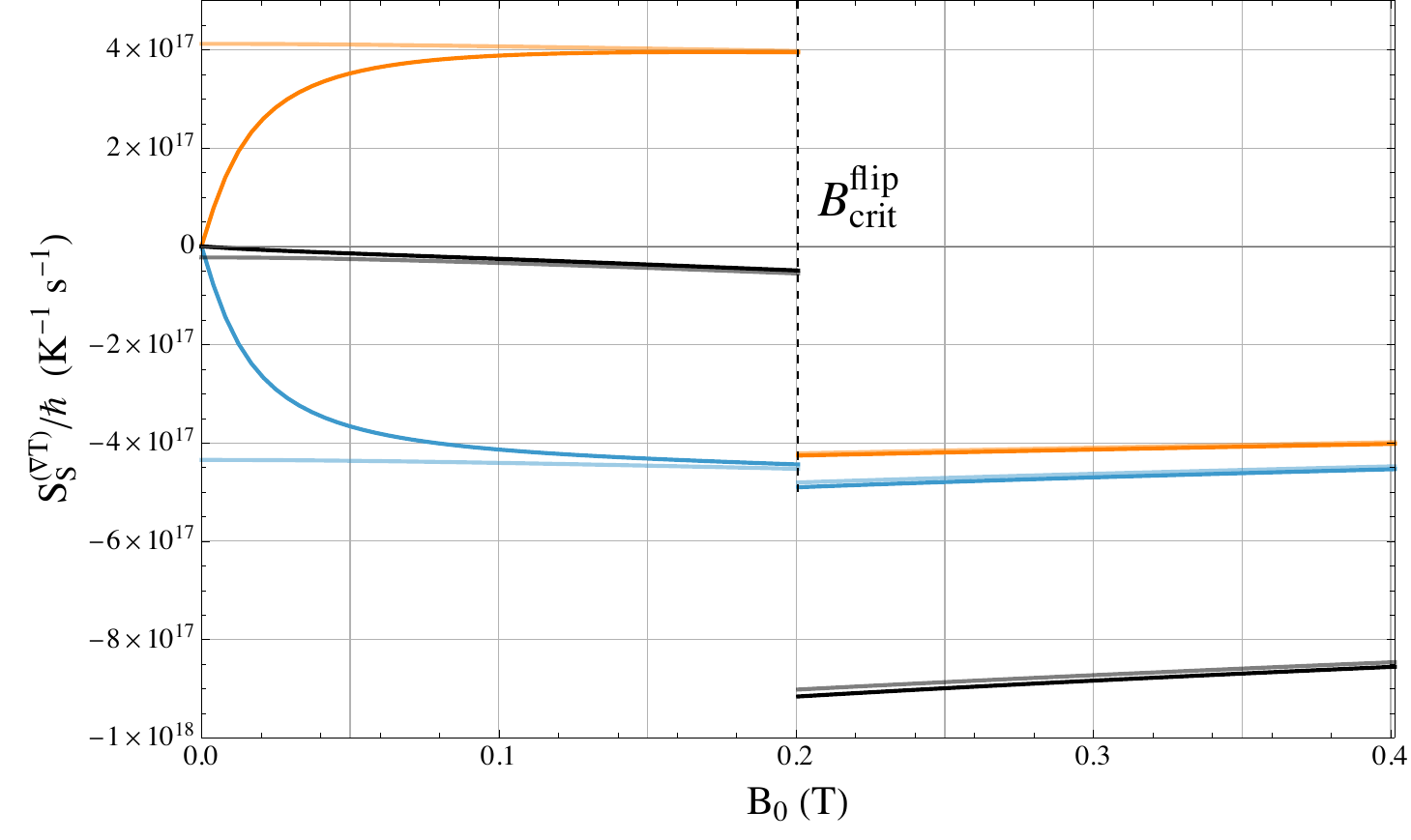}%
%}
\label{fig5a}

%\subfloat[]{%
  \hspace{-3mm}\includegraphics[clip,width=0.985\columnwidth]{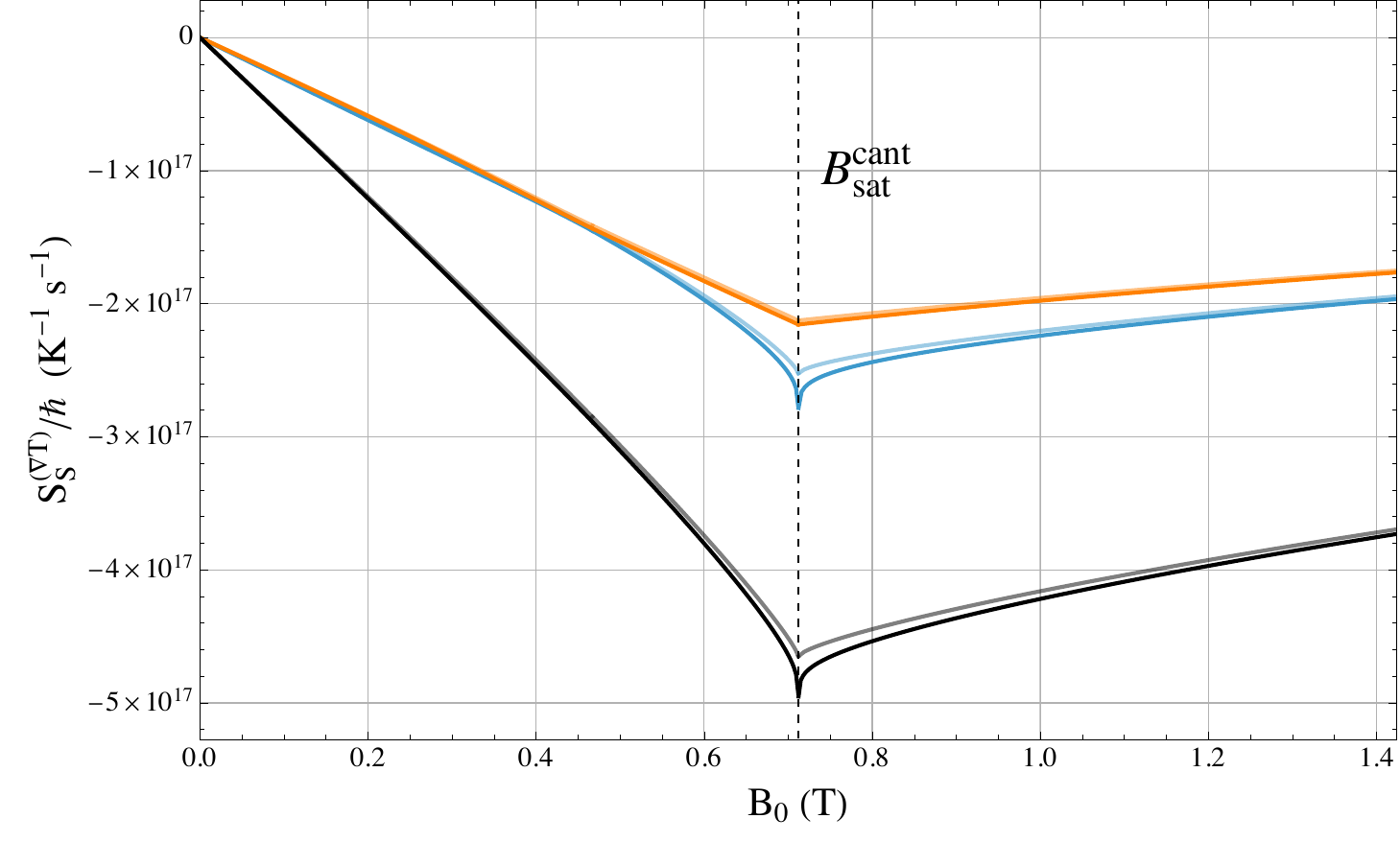}%
%}
\label{fig5b}

\caption{ \small Thermal SSC of the two magnon modes and their sum (black curves) in the bilayer as a function of the external magnetic field \(B_0\). The semi-transparent curves correspond to \(S_S^{(\nabla T)}\) with the SAM set to \(\pm \hbar\) (\(\hbar\)) in the AFM (FM) phase. (a) \(S_S^{(\nabla T)}\) along the easy axis for $\vec{B}_0 \parallel \hat{b}$ in the AFM (\(B_0 < B^{\text{flip}}_{\text{crit}}\)) and FM (\(B_0 \geq B^{\text{flip}}_{\text{crit}}\)) phases. (b) \(S_S^{(\nabla T)}\) along the intermediate axis for $\vec{B}_0 \parallel \hat{a}$ in the canted (\(B_0 < B^{\text{cant}}_{\text{sat}}\)) and saturated (\(B_0 \geq B^{\text{cant}}_{\text{sat}}\)) phases.}
\end{figure}

Fig. \hyperlink{fig6a}{6(a)} shows the diffusive SSC of the bilayer in the AFM and FM phases at \(T=5\) K. In both phases the diffusive SSC behaves much like the thermal one. For \(B_0<B_{\mathrm{crit}}^{\mathrm{flip}}\) the two magnon branches carry opposite SAM. Naively setting the SAM to \(\pm \hbar\) overestimates the diffusive SSC which does not accurately go to zero for external fields near zero. Above the spin-flip transition both branches carry spin with the same sign and show a weak field dependence.

Fig. \hyperlink{fig6b}{6(b)} shows the diffusive SSC of the bilayer in the canted and saturated phases at \(T=5\) K. For \(\vec{B}_0 \parallel \hat{a}\), the lower magnon mode softens as the external field approaches \(B_{\mathrm{sat}}^{\mathrm{cant}}\), producing a pronounced enhancement of the diffusive SSC peak. As in the monolayer, in contrast to the thermal SSC, the diffusive SSC exhibits a peak for the soft mode even when the SAM is fixed to $\Delta S=\hbar$ (semi-transparent curves). The non-universal SAM further amplifies this low-energy contribution. The relatively hard mode remains weakly field-dependent.

Figs. \hyperref[fig15]{15} and \hyperref[fig17]{17} show the total diffusive SSC and magnon thermal conductivity, respectively, for various temperatures in all phases in the bilayer.

\begin{figure} [ht] \label{fig6}

%\subfloat[]{%
  \hspace{-3mm}\includegraphics[clip,width=1\columnwidth]{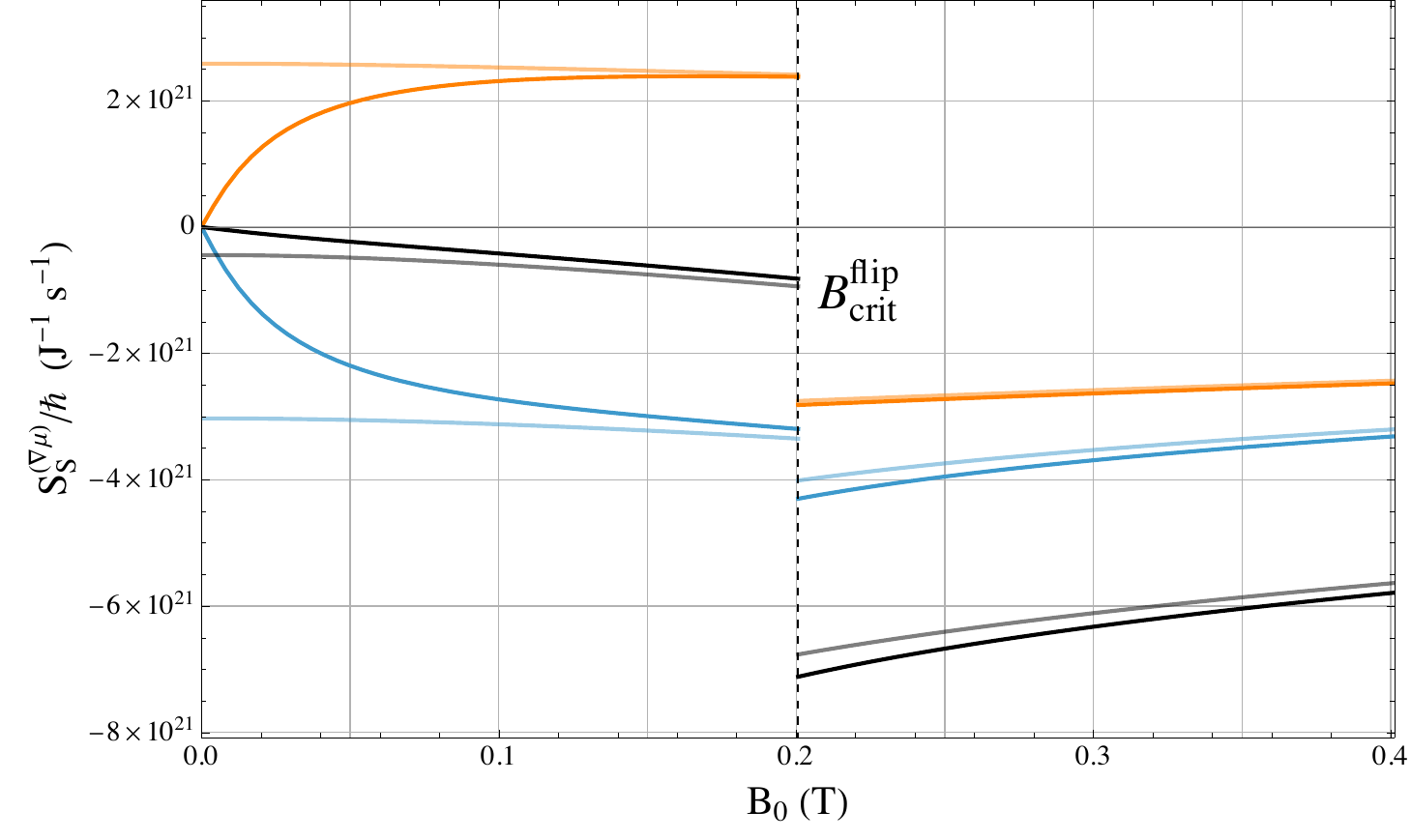}%
%}
\label{fig6a}

%\subfloat[]{%
  \hspace{-3mm}\includegraphics[clip,width=0.985\columnwidth]{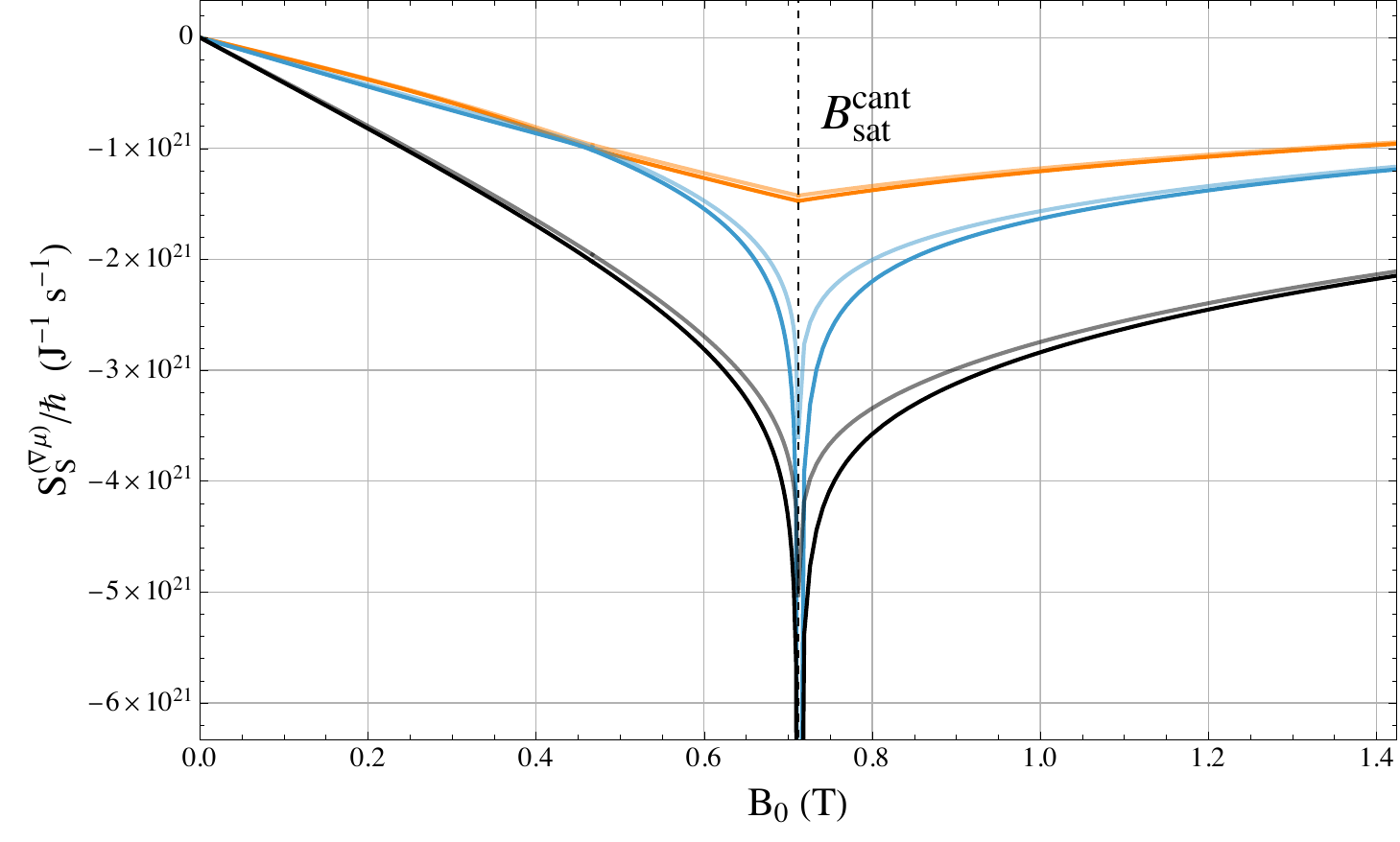}%
%}
\label{fig6b}

\caption{ \small Diffusive SSC of the two magnon modes and their sum (black curves) in the bilayer as a function of the external magnetic field \(B_0\). The semi-transparent curves correspond to \(S_S^{(\nabla \mu)}\) with the SAM set to \(\pm \hbar\) (\(\hbar\)) in the AFM (FM) phase. (a) \(S_S^{(\nabla \mu)}\) along the easy axis for $\vec{B}_0 \parallel \hat{b}$ in the AFM (\(B_0 < B^{\text{flip}}_{\text{crit}}\)) and FM (\(B_0 \geq B^{\text{flip}}_{\text{crit}}\)) phases. (b) \(S_S^{(\nabla \mu)}\) along the intermediate axis for $\vec{B}_0 \parallel \hat{a}$ in the canted (\(B_0 < B^{\text{cant}}_{\text{sat}}\)) and saturated (\(B_0 \geq B^{\text{cant}}_{\text{sat}}\)) phases.}
\end{figure}

\section{\label{V}Conclusion}
We calculated the non-universal magnon SAM and the associated thermal and diffusive SSC for mono- and bilayer CrSBr with intralayer triaxial anisotropy and intralayer dipolar interactions, a particularly suitable platform with magnon modes in the GHz regime. The frequency of one of the two magnon modes in the antiferromagnet softens to zero under moderate in-plane magnetic fields perpendicular to the zero-field Neel vector, which leads to an algebraic divergence of the SAM and a logarithmic one in the thermal SSC. The latter allows for an experimentally accessible spin-caloritronic signature of soft magnons. In the diffusive SSC, the peak arising from the strong sensitivity of the response to the chemical potential at low frequencies is further enhanced by the non-universal SAM.

While our analysis focuses on bulk magnon transport, experiments are usually carried out electrically with heavy metal contacts to the vdW magnet. In contrast to Pt contacts on yttrium iron garnet films, the interface conductance that depends strongly on the degree of compensation \cite{Tang2024} may play an important role. 

Our results highlight the effect of non-universal magnon SAM in spin caloritronics. Future work should address how these bulk spin-caloritronic signatures are modified by interfacial effects, such as spin-mixing conductance \cite{Heinrich2011,Tang2024}, magnon-electron conversion efficiency \cite{Zhang2012}, and spin backflow at metal-magnetic insulator interfaces \cite{Lu2019}. Experiments on CrSBr and other vdW magnets are necessary to assess the relative importance of these contributions. 

\begin{acknowledgments}
This publication is part of the project "Ronde Open Competitie XL" (file number OCENW.XL21.XL21.058) which is financed by the Dutch Research Council (NWO). G.B. and P.T. were supported by JSPS KAKENHI Grants (No. 22H04965 and No. 24H02231). Images were made with BioRender and the crystallographic model was taken from VESTA.
\end{acknowledgments}

\section*{Data Availablity}
The data supporting the findings of this article are openly
available \cite{DataTeuling}.

\appendix

\section{CrSBr model and material values} \label{appA}

\begin{figure}[ht] \label{fig7}
    \centering
    \includegraphics[scale=0.35]{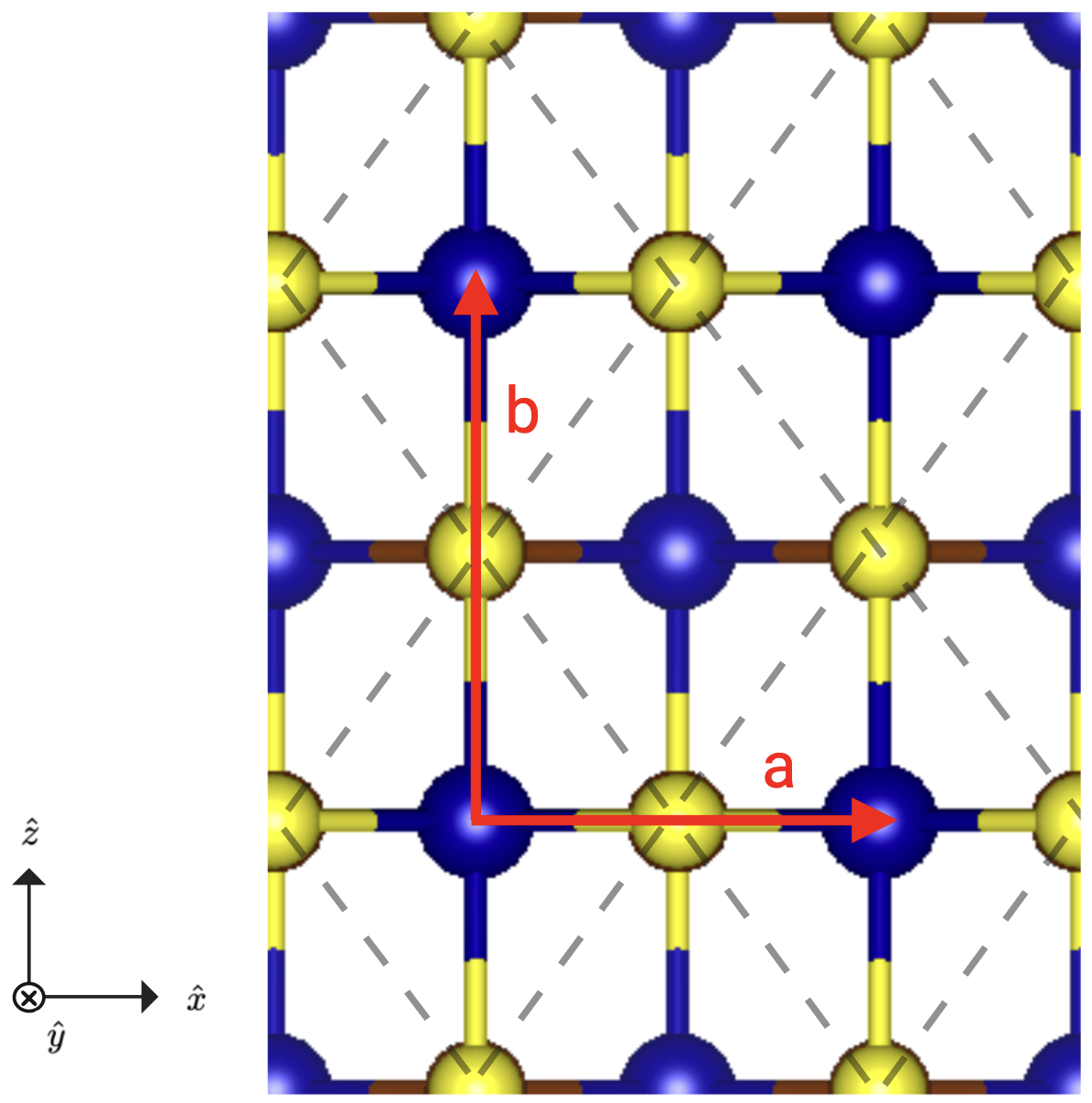}
    \caption{  \small Top view of monolayer CrSBr illustrating the crystallographic axes and lattice vectors. Primitive unit cell in dashed lines. The magnetic chromium  atoms are colored blue.}
\end{figure}

Fig. \hyperref[fig7]{7} shows the CrSBr crystal structure and Table \hyperref[tab1]{I} lists the parameters used here.

\begin{table}[ht]
\caption{\label{tab1}%
In-plane lattice parameters \(a\) and \(b\), exchange interactions \(\mathcal{J}_i\) \cite{Esteras2022}, and anisotropy coefficients \(\mathcal{D}_i\) \cite{Yang2021}.
}
\begin{ruledtabular}
\begin{tabular}{ccc}
    $\emph{a}$ & 3.54 & $10^{-10} \mathrm{m}$ \\ 
    $\emph{b}$ & 4.73 & $10^{-10}\mathrm{m}$ \\ \hline
    $\mathcal{J}_1$ & 3.54 & $\mathrm{meV}$ \\ 
    $\mathcal{J}_2$ & 3.08 & $\mathrm{meV}$ \\
    $\mathcal{J}_3$ & 4.15 & $\mathrm{meV}$ \\ 
    $\mathcal{J}_{\perp}$ & -15.5 & $\mu$$\mathrm{eV}$ \\ \hline 
   $ \mathcal{D}_x$ & -12 & $\mu$$\mathrm{eV}$ \\
    $\mathcal{D}_y$ & -78 & $\mu$$\mathrm{eV}$ \\
    $\mathcal{D}_z$ & 0 & $\mu$$\mathrm{eV}$ \\ 
\end{tabular}
\end{ruledtabular}
\end{table}

\section{Monolayer} \label{appB}

\subsection{Hamiltonian}\label{appB1}

The Hamiltonian for a monolayer reads \cite{Teuling2}
\begin{equation} \label{B1}
    \begin{aligned}
        {\hspace{-0.5mm}}H_A = & \underbrace{-\sum_{j}\gamma \hbar \vec{B}_0\cdot{\hspace{-0.5mm}}\vec{S}_{j,A}}_{H_{Ext}} \underbrace{-\sum_{j,\sigma} \mathcal{J}_{\sigma}{\hspace{-0.5mm}}\vec{S}_{j,A} \cdot {\hspace{-0.5mm}}\vec{S}_{j+\sigma,A}}_{H_{Ex}} \\
        & \underbrace{-\sum_{j} \Big[\mathcal{D}_x\hspace{1mm} {\hspace{-0.5mm}}S_{x,j,A}^2 + \mathcal{D}_y\hspace{1mm} {\hspace{-0.5mm}}S_{y,j,A}^2 + \mathcal{D}_z\hspace{1mm} {\hspace{-0.5mm}}S_{z,j,A}^2 \Big]}_{H_{An}}, \\
    \end{aligned}
\end{equation}

\noindent
to which we add \(H_{dip,A}\) from Eq. (\hyperref[2]{2}).

\subsection{Resonance frequencies}\label{appB2}

\begin{figure}[ht] \label{fig8}
    \centering
    \includegraphics[scale=0.35]{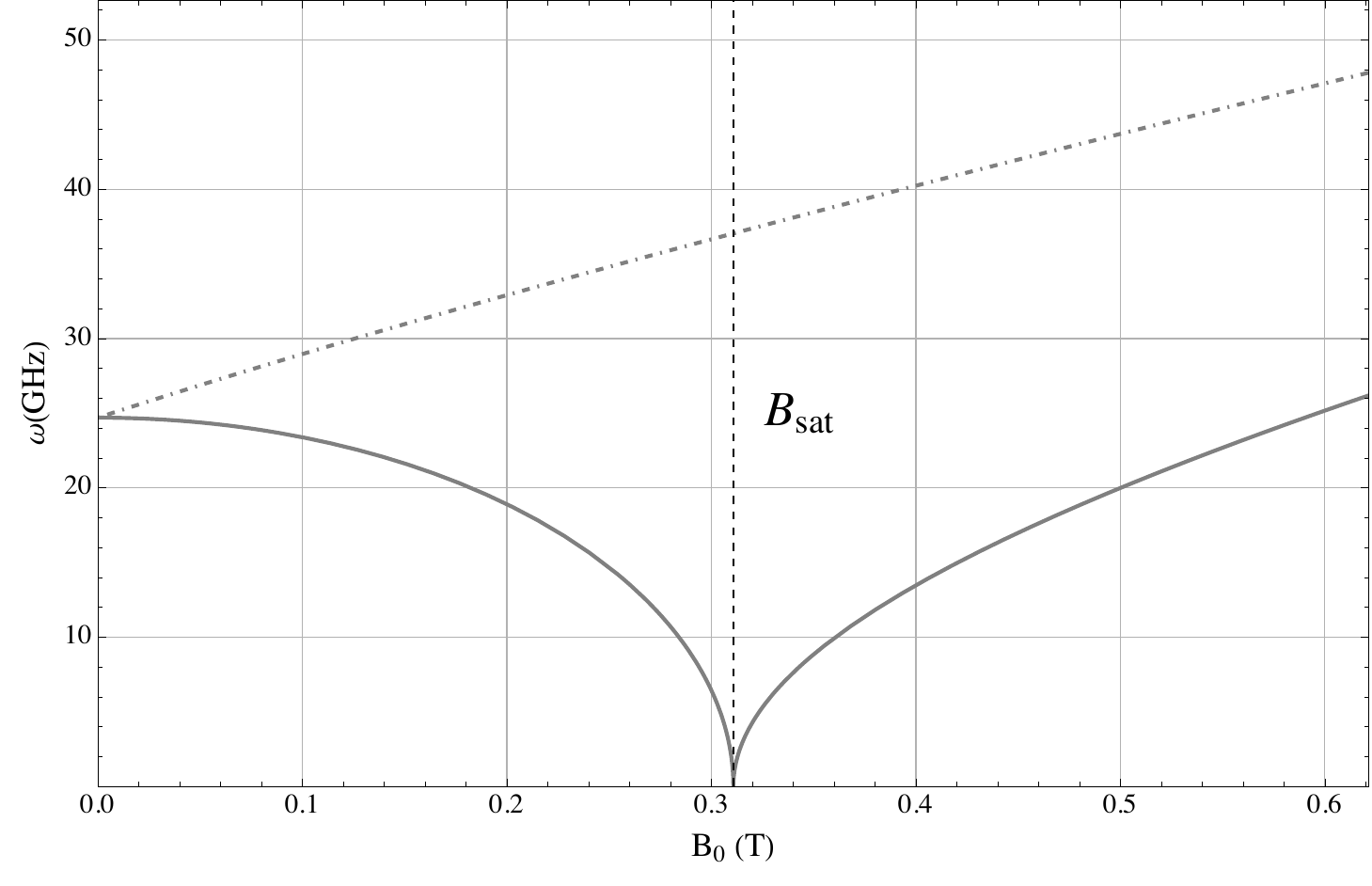}
    \caption{  \small Resonance frequencies (\(k=0\)) in the monolayer. The dash-dotted curve corresponds to  \( \vec{B}_0\parallel \hat{z} \) (\(\hat{b}\)), the solid curve corresponds to  \( \vec{B}_0\parallel \hat{x} \) (\(\hat{a}\)).}
\end{figure}

The resonance frequencies for all phases in the monolayer have been calculated in Ref. \cite{Teuling2} and are shown in Fig. \hyperref[fig8]{8}.

\subsection{Canted configuration}\label{appB3}

\begin{figure}[ht] \label{fig9}
    \centering
    \includegraphics[scale=0.35]{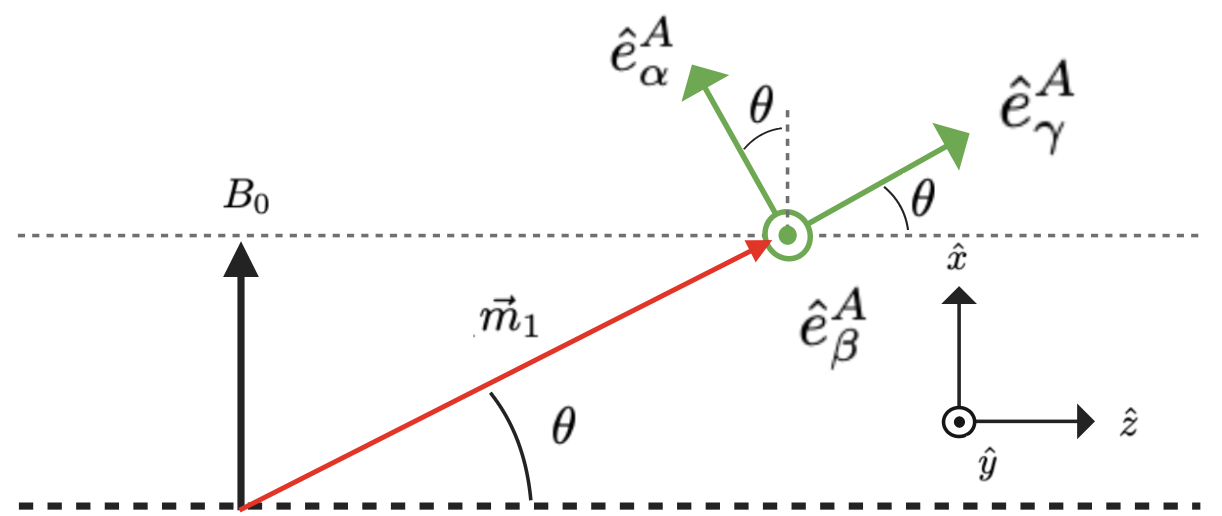}
    \caption{  \small Canted configuration of magnetic moments in the monolayer for \( \vec{B}_0\parallel \hat{x} \) (\(\hat{a}\)), showing the local rotated reference frame and the corresponding unit vectors.}
\end{figure}

Here we describe the coordinate transformations when the spin texture is non-collinear to an anisotropy axis. The canted phase in the monolayer is described in the basis of \([\hat{e}^A_{\alpha},\hat{e}^A_{\beta},\hat{e}^A_{\gamma}]\) with transformations 
\begin{equation} \label{B2}
    \begin{aligned}
        & \hat{x} =  \cos (\theta)\hat{e}^A_{\alpha}  + \sin (\theta)\hat{e}^A_{\gamma}  , \\
        & \hat{y} = \hat{e}^A_{\beta}, \\
        & \hat{z} = -\sin(\theta)\hat{e}^A_{\alpha} + \cos(\theta)\hat{e}^A_{\gamma},
    \end{aligned}
\end{equation}

\noindent
as illustrated in Fig. \hyperref[fig9]{9}.

\subsection{Eigenvalue matrices}\label{appB4}

Here we show the matrix elements of \(\mathcal{H}^{(l)}\) from Eq. (\hyperref[4]{4}) for all phases in the monolayer. The elements for the external field along the easy axis (\(b\)) in the monolayer read
\begin{equation} \label{B3}
\begin{aligned}
    A^{(f)}_k & = 2S\gamma(k) + S(\mathcal{D}_x + \mathcal{D}_y) - \hbar \gamma B_0 - 2S(\mathcal{J} + \mathcal{D}_z)\\
    & - \frac{1}{2}\mu_0 \gamma \hbar M_s f(k) - \frac{1}{2}\mu_0 \gamma \hbar M_s \frac{k_{x}^2}{k^2}(1-f(k)), \\
    B^{(f)}_k & = S(\mathcal{D}_x - \mathcal{D}_y) \\
    &+ \frac{1}{2}\mu_0 \gamma \hbar M_s f(k)- \frac{1}{2}\mu_0 \gamma \hbar M_s \frac{k_{x}^2}{k^2}(1-f(k)).\\
\end{aligned}
\end{equation}

The elements for the canted phase in the monolayer read
\begin{equation} \label{B4}
\begin{aligned}
    A^{m,c}_k & = 2S\gamma(k) + S(\mathcal{D}_x \cos^2(\theta) + \mathcal{D}_z \sin^2(\theta) + \mathcal{D}_y) \\
    & - \hbar \gamma B_0\sin(\theta) - 2S(\mathcal{J} + \mathcal{D}_x \sin^2(\theta) + \mathcal{D}_z \cos^2(\theta))\\
    & - \frac{1}{2}\mu_0 \gamma \hbar M_s f(k)- \frac{1}{2}\mu_0 \gamma \hbar M_s \frac{k_{\alpha}^2}{k^2}(1-f(k)), \\
    B^{m,c}_k & = S(\mathcal{D}_x \cos^2(\theta) + \mathcal{D}_z \sin^2(\theta) - \mathcal{D}_y) \\
    &+ \frac{1}{2}\mu_0 \gamma \hbar M_s f(k) - \frac{1}{2}\mu_0 \gamma \hbar M_s \frac{k_{\alpha}^2}{k^2}(1-f(k)).\\
\end{aligned}
\end{equation}

The elements for the saturated phase in the monolayer read
\begin{equation} \label{B5}
\begin{aligned}
    A^{m,s}_k & = 2S\gamma(k) + S(\mathcal{D}_z + \mathcal{D}_y) - \hbar \gamma B_0 - 2S(\mathcal{J} + \mathcal{D}_x)\\
    & - \frac{1}{2}\mu_0 \gamma \hbar M_s f(k)- \frac{1}{2}\mu_0 \gamma \hbar M_s \frac{k_{z}^2}{k^2}(1-f(k)), \\
    B^{m,s}_k & = S(\mathcal{D}_z - \mathcal{D}_y) \\
    &+ \frac{1}{2}\mu_0 \gamma \hbar M_s f(k)- \frac{1}{2}\mu_0 \gamma \hbar M_s \frac{k_{z}^2}{k^2}(1-f(k)).\\
\end{aligned}
\end{equation}

\noindent
\(k_{\alpha,i}\) (\(i=A,B\)) is related to \(k_x\) and \(k_z\) through Eq. (\hyperlink{B2}{B2}). In all phases
\begin{equation} \label{B6}
\begin{aligned}
    \gamma(k) &= 2\mathcal{J}_1\cos(k_xa) + 2\mathcal{J}_3\cos(k_zb)\\
    & + 2\mathcal{J}_2(\cos(\frac{1}{2}k_xa + \frac{1}{2} k_zb) + \cos(\frac{1}{2}k_xa - \frac{1}{2} k_zb)),
\end{aligned}
\end{equation}

\noindent
and
\begin{equation} \label{B7}
    \mathcal{J} = 2\mathcal{J}_1 + 4\mathcal{J}_2 + 2\mathcal{J}_3.
\end{equation}

\section{Bilayer} \label{appC}

\subsection{Hamiltonian} \label{appC1}

The Hamiltonian for a bilayer reads \cite{Teuling2}
\begin{equation} \label{C1}
    H_{Bi} = {\hspace{-0.5mm}}H_A +{\hspace{-0.5mm}}H_B + H_{int}.
\end{equation}

\noindent
where

\begin{equation} \label{C2}
    H_{int} = -\sum_{j} \mathcal{J}_{\perp}\hspace{-0.5mm}\vec{S}_{j,A} \cdot \hspace{-0.5mm}\vec{S}_{j,B},
\end{equation}

\noindent
\noindent
to which we add \(H_{dip,A}\) and \(H_{dip,B}\) from Eq. (\hyperref[2]{2}).

\subsection{(A)FMR} \label{appC2}

\begin{figure} [ht] \label{fig10}

%\subfloat[]{%
  \hspace{-3mm}\includegraphics[clip,width=1\columnwidth]{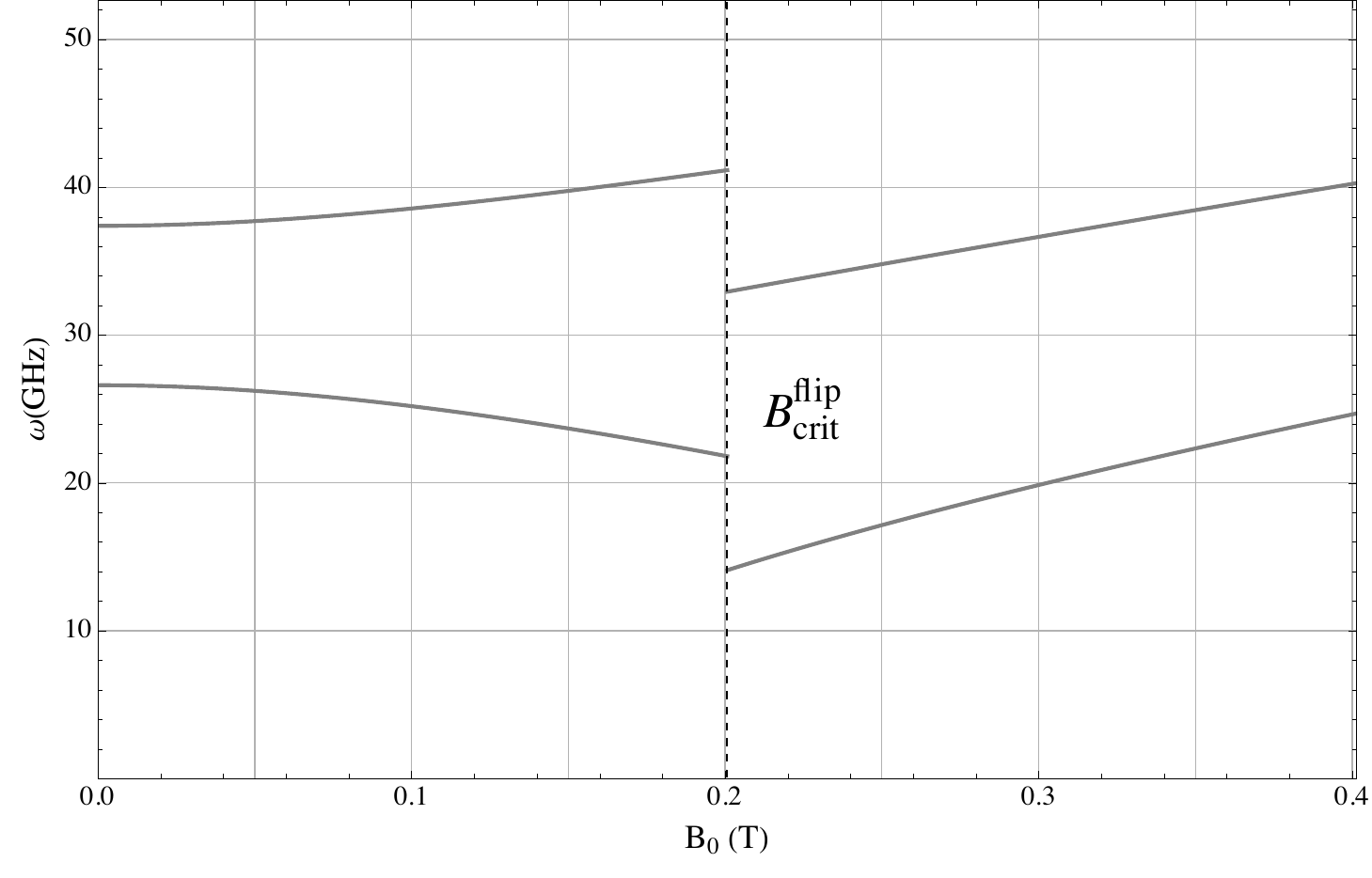}%
%}
\label{fig10a}

%\subfloat[]{%
  \hspace{-3mm}\includegraphics[clip,width=1\columnwidth]{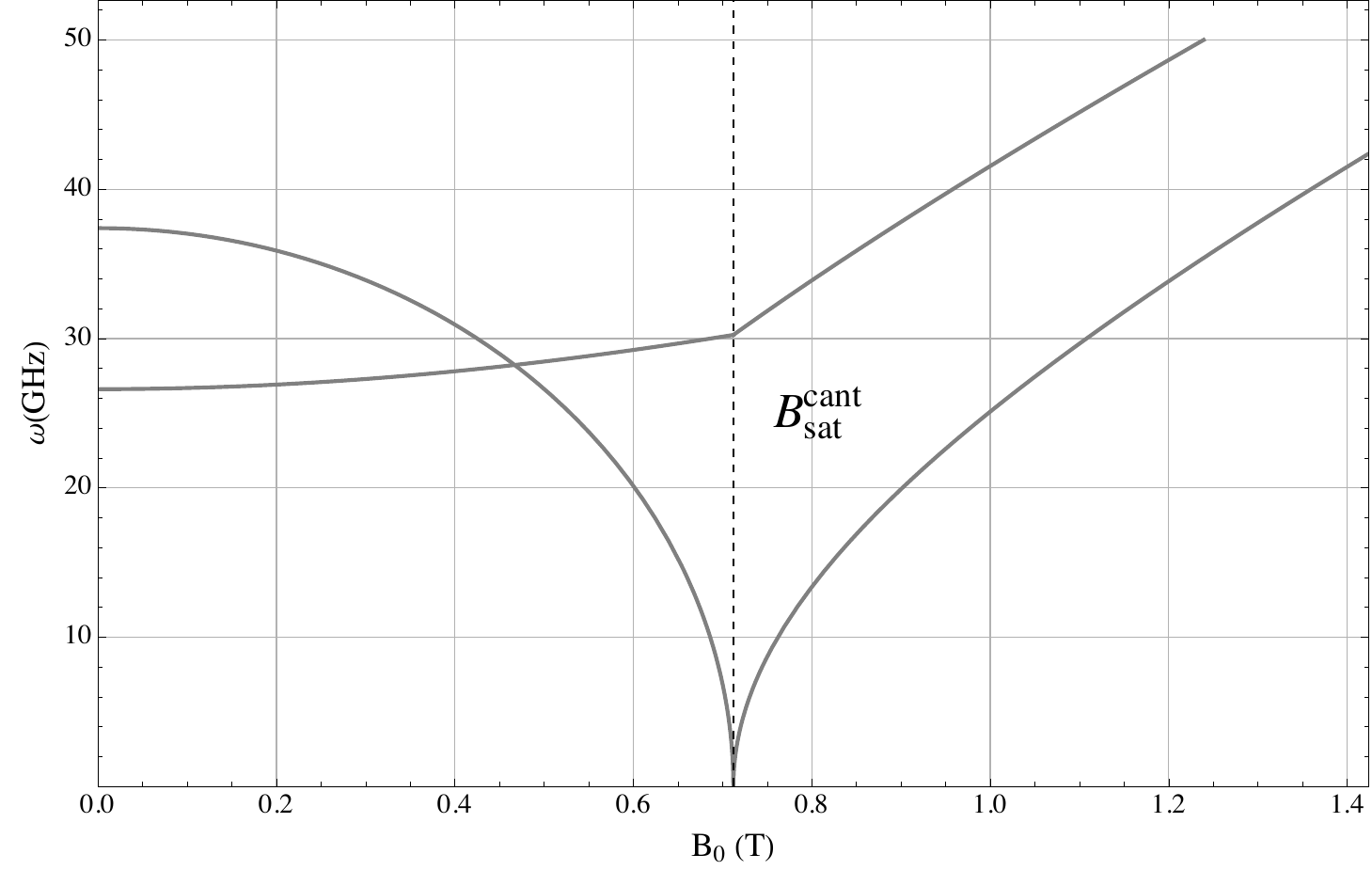}%
%}
\label{fig10b}

\caption{ \small Resonance frequencies (\(k=0\)) in the bilayer. (a) AFM (\(\vec{B}_0 < B^{flip}_{crit}\)) and FM (\(\vec{B}_0 \geq B^{flip}_{crit}\)) phases with \( \vec{B}_0\parallel \hat{z} \) (\(\hat{b}\)). (b) Canted (\(\vec{B}_0 < B^{cant}_{sat}\)) and saturated (\(\vec{B}_0 \geq B^{cant}_{sat}\)) phases with \( \vec{B}_0\parallel \hat{x} \) (\(\hat{a}\)).}
\end{figure}

The resonance frequencies for all phases in the bilayer have been calculated in Ref. \cite{Teuling2} and are shown in Fig. \hyperref[fig10]{10}.

\subsection{Canted configuration} \label{appC3}

\begin{figure}[ht] \label{fig11}
    \centering
    \includegraphics[scale=0.295]{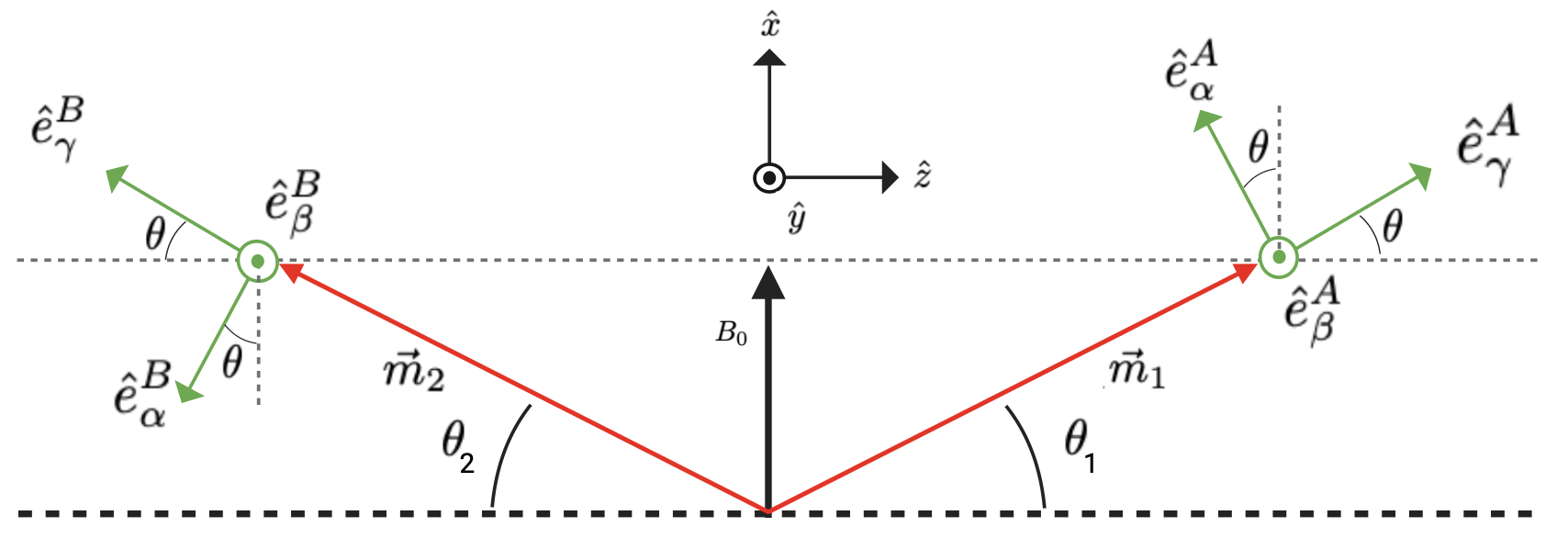}
    \caption{  \small Canted configuration of magnetic moments in the bilayer for \( \vec{B}_0\parallel \hat{x} \) (\(\hat{a}\)), showing the local rotated reference frames and the corresponding unit vectors.}
\end{figure}

Here we describe the coordinate transformations when the spin texture is non-collinear to an anisotropy axis. The canted phase in the bilayer is described in the basis of 
\(
[\hat{e}^A_{\alpha},\hat{e}^A_{\beta},\hat{e}^A_{\gamma}]\) and \([\hat{e}^B_{\alpha},\hat{e}^B_{\beta},\hat{e}^B_{\gamma}]\) with transformations
\begin{equation} \label{C3}
    \begin{aligned}
        & \hat{x} = \cos(\theta)\hat{e}^A_{\alpha} + \sin(\theta)\hat{e}^A_{\gamma}, \\
        & \hat{y} = \hat{e}^A_{\beta}, \\
        & \hat{z} = -\sin(\theta)\hat{e}^A_{\alpha} + \cos(\theta)\hat{e}^A_{\gamma},
    \end{aligned}
\end{equation}

\noindent
and
\begin{equation} \label{C4}
    \begin{aligned}
        & \hat{x} = -\cos(\theta)\hat{e}^B_{\alpha} + \sin(\theta)\hat{e}^B_{\gamma}, \\
        & \hat{y} = \hat{e}^B_{\beta}, \\
        & \hat{z} = -\sin(\theta)\hat{e}^B_{\alpha} - \cos(\theta)\hat{e}^B_{\gamma},
    \end{aligned}
\end{equation}

\noindent
as illustrated in Fig. \hyperref[fig11]{11}.

\subsection{Eigenvalue matrix} \label{appC4}

Here we show the matrix \(\mathcal{H}^{(l)}\) for all phases in the bilayer. The matrix for the AFM phase in the bilayer reads
\begin{equation} \label{C5}
\begin{aligned}
\mathcal{H}^{AFM}=
\begin{bmatrix}
-  A^{A,\parallel z}_k & 0 & - B^{A,\parallel z}_k &  \mathcal{J}_{\perp} S \\
0 & -  A^{B,\parallel -z}_k  &  \mathcal{J}_{\perp} S & - B^{A,\parallel z}_k \\
- B^{A,\parallel z}_k &  \mathcal{J}_{\perp} S & -  A^{A,\parallel z}_k & 0 \\
 \mathcal{J}_{\perp} S & - B^{A,\parallel z}_k & 0 & - A^{B,\parallel -z}_k
\end{bmatrix},
\end{aligned}
\end{equation}

\noindent
where
\begin{equation} \label{C6}
\begin{aligned}
    A^{A,\parallel z}_k & = 2S\gamma(k) + S(\mathcal{D}_x + \mathcal{D}_y) - \hbar \gamma B_0 - 2S(\mathcal{J} + \mathcal{D}_z)\\
    & + S\mathcal{J}_{\perp} - \frac{1}{2}\mu_0 \gamma \hbar M_s f(k) - \frac{1}{2}\mu_0 \gamma \hbar M_s \frac{k_{x}^2}{k^2}(1-f(k)),\\
    A^{B,\parallel -z}_k & = 2S\gamma(k) + S(\mathcal{D}_x + \mathcal{D}_y) + \hbar \gamma B_0 - 2S(\mathcal{J} + \mathcal{D}_z)\\
    & + S\mathcal{J}_{\perp}- \frac{1}{2}\mu_0 \gamma \hbar M_s f(k) - \frac{1}{2}\mu_0 \gamma \hbar M_s \frac{k_{x}^2}{k^2}(1-f(k))\\
    &+ \frac{1}{2}\mu_0 \gamma \hbar M_s f(k)- \frac{1}{2}\mu_0 \gamma \hbar M_s \frac{k_{x}^2}{k^2}(1-f(k)),\\
    B^{A,\parallel z}_k & = S(\mathcal{D}_x - \mathcal{D}_y)\\
    &+ \frac{1}{2}\mu_0 \gamma \hbar M_s f(k)- \frac{1}{2}\mu_0 \gamma \hbar M_s \frac{k_{x}^2}{k^2}(1-f(k)).\\
\end{aligned}
\end{equation}

\noindent
The superscripts indicate the magnetization of sublattice \(A\) (\(B\)) is along \(z\) (\(-z\)).

The matrix for the FM phase in the bilayer reads
\begin{equation} \label{C7}
    \mathcal{H}^{FM} = 
\begin{bmatrix}
-A_k^{FM} & -\mathcal{J}_\perp S & -B_k^{FM} & 0 \\
-\mathcal{J}_\perp S & -A_k^{FM} & 0 & -B_k^{FM} \\
-B_k^{FM} & 0 & -A_k^{FM} & -\mathcal{J}_\perp S \\
0 & -B_k^{FM} & -\mathcal{J}_\perp S & -A_k^{FM}
\end{bmatrix},
\end{equation}

\noindent
where
\begin{equation} \label{C8}
\begin{aligned}
    A^{FM}_k & = 2S\gamma(k) + S(\mathcal{D}_x + \mathcal{D}_y) - \hbar \gamma B_0 \\
    & - 2S(\mathcal{J} + \mathcal{D}_z) -S\mathcal{J}_{\perp}\\
    & - \frac{1}{2}\mu_0 \gamma \hbar M_s f(k) - \frac{1}{2}\mu_0 \gamma \hbar M_s \frac{k_{x}^2}{k^2}(1-f(k)), \\
    B^{FM}_k & = S(\mathcal{D}_x - \mathcal{D}_y)\\
    &+ \frac{1}{2}\mu_0 \gamma \hbar M_s f(k)- \frac{1}{2}\mu_0 \gamma \hbar M_s \frac{k_{x}^2}{k^2}(1-f(k)).\\
\end{aligned}
\end{equation}

The matrix for the canted phase in the bilayer reads
\begin{equation} \label{C9}
\mathcal{H}^{b,c} = 
\begin{bmatrix}
- A_k^{A,\parallel \gamma^A} & -M^+ & -B_k^{A,\parallel \gamma^A} & -M^- \\
-M^+ & - A_k^{B,\parallel \gamma^B} & -M^- & -B_k^{B,\parallel \gamma^B} \\
- B_k^{A,\parallel \gamma^A} & -M^- & - A_k^{A,\parallel \gamma^A} & -M^+ \\
- M^- & - B_k^{B,\parallel \gamma^B} & -M^+ & - A_k^{B,\parallel \gamma^B}
\end{bmatrix},
\end{equation}

\noindent
where
\begin{equation} \label{C10}
\begin{aligned}
    A^{A,\parallel \gamma^A}_k & = 2S\gamma(k) + S(\mathcal{D}_x \cos^2(\theta) + \mathcal{D}_z \sin^2(\theta) + \mathcal{D}_y) \\
    & - \hbar \gamma B_0\sin(\theta) - 2S(\mathcal{J} + \mathcal{D}_x \sin^2(\theta) + \mathcal{D}_z \cos^2(\theta))\\
    & + \cos(2\theta)S\mathcal{J}_{\perp}\\
    & - \frac{1}{2}\mu_0 \gamma \hbar M_s f(k)- \frac{1}{2}\mu_0 \gamma \hbar M_s \frac{k_{\alpha,A}^2}{k^2}(1-f(k)), \\
    A^{B,\parallel \gamma^B}_k & = 2S\gamma(k) + S(\mathcal{D}_x \cos^2(\theta) + \mathcal{D}_z \sin^2(\theta) + \mathcal{D}_y) \\
    & - \hbar \gamma B_0\sin(\theta) - 2S(\mathcal{J} + \mathcal{D}_x \sin^2(\theta) + \mathcal{D}_z \cos^2(\theta))\\
    & + \cos(2\theta)S\mathcal{J}_{\perp}\\
    & - \frac{1}{2}\mu_0 \gamma \hbar M_s f(k)- \frac{1}{2}\mu_0 \gamma \hbar M_s \frac{k_{\alpha,B}^2}{k^2}(1-f(k)), \\
    B^{A,\parallel \gamma^A}_k & = S(\mathcal{D}_x \cos^2(\theta) + \mathcal{D}_z \sin^2(\theta) - \mathcal{D}_y)\\
    &+ \frac{1}{2}\mu_0 \gamma \hbar M_s f(k) - \frac{1}{2}\mu_0 \gamma \hbar M_s \frac{k_{\alpha,A}^2}{k^2}(1-f(k)),\\
    B^{B,\parallel \gamma^B}_k & = S(\mathcal{D}_x \cos^2(\theta) + \mathcal{D}_z \sin^2(\theta) - \mathcal{D}_y)\\
    &+ \frac{1}{2}\mu_0 \gamma \hbar M_s f(k) - \frac{1}{2}\mu_0 \gamma \hbar M_s \frac{k_{\alpha,B}^2}{k^2}(1-f(k)),\\
    M^+ &= \frac{1}{2}\mathcal{J}_{\perp}S(1-\cos(2\theta)),\\
    M^- & = -\frac{1}{2}\mathcal{J}_{\perp}S(1+\cos(2\theta)).\\
\end{aligned}
\end{equation}

\noindent
The superscripts indicate the magnetization of sublattice \(A\) (\(B\)) is along \(\hat{e}^A_{\gamma}\) (\(\hat{e}^B_{\gamma}\)) as shown in Fig. \hyperref[fig11]{11}.

The matrix for the saturated phase in the bilayer reads
\begin{equation} \label{C11}
    \mathcal{H}^{b,s} = 
\begin{bmatrix}
-A_k^{b,s} & -\mathcal{J}_\perp S & -B_k^{b,s} & 0 \\
-\mathcal{J}_\perp S & -A_k^{b,s} & 0 & -B_k^{b,s} \\
-B_k^{b,s} & 0 & -A_k^{b,s} & -\mathcal{J}_\perp S \\
0 & -B_k^{b,s} & -\mathcal{J}_\perp S & -A_k^{b,s}
\end{bmatrix},
\end{equation}

\noindent
where
\begin{equation} \label{C12}
\begin{aligned}
    A^{b,s}_k & = 2S\gamma(k) + S(\mathcal{D}_z + \mathcal{D}_y) - \hbar \gamma B_0 \\
    & - 2S(\mathcal{J} + \mathcal{D}_x) -S\mathcal{J}_{\perp}\\
    & - \frac{1}{2}\mu_0 \gamma \hbar M_s f(k) - \frac{1}{2}\mu_0 \gamma \hbar M_s \frac{k_{z}^2}{k^2}(1-f(k)), \\
    B^{b,s}_k & = S(\mathcal{D}_z - \mathcal{D}_y)\\
    &+ \frac{1}{2}\mu_0 \gamma \hbar M_s f(k)- \frac{1}{2}\mu_0 \gamma \hbar M_s \frac{k_{z}^2}{k^2}(1-f(k)).\\
\end{aligned}
\end{equation}

\noindent
and \(k_{\alpha,i}\) (\(i=A,B\)) is related to \(k_x\) and \(k_z\) through Eqs. (\hyperlink{C3}{C3}) and (\hyperlink{C4}{C4}). In all phases
\begin{equation} \label{C13}
\begin{aligned}
    \gamma(k) &= 2\mathcal{J}_1\cos(k_xa) + 2\mathcal{J}_3\cos(k_zb)\\
    & + 2\mathcal{J}_2(\cos(\frac{1}{2}k_xa + \frac{1}{2} k_zb) + \cos(\frac{1}{2}k_xa - \frac{1}{2} k_zb)),
\end{aligned}
\end{equation}

\noindent
and
\begin{equation} \label{C14}
    \mathcal{J} = 2\mathcal{J}_1 + 4\mathcal{J}_2 + 2\mathcal{J}_3.
\end{equation}

\section{SSC at various temperatures} \label{appD}

\subsection{Monolayer} \label{appD1}

\begin{figure}[ht] \label{fig12}
  %\subfloat{%
    \hspace{-3mm}\includegraphics[clip,width=1\columnwidth]{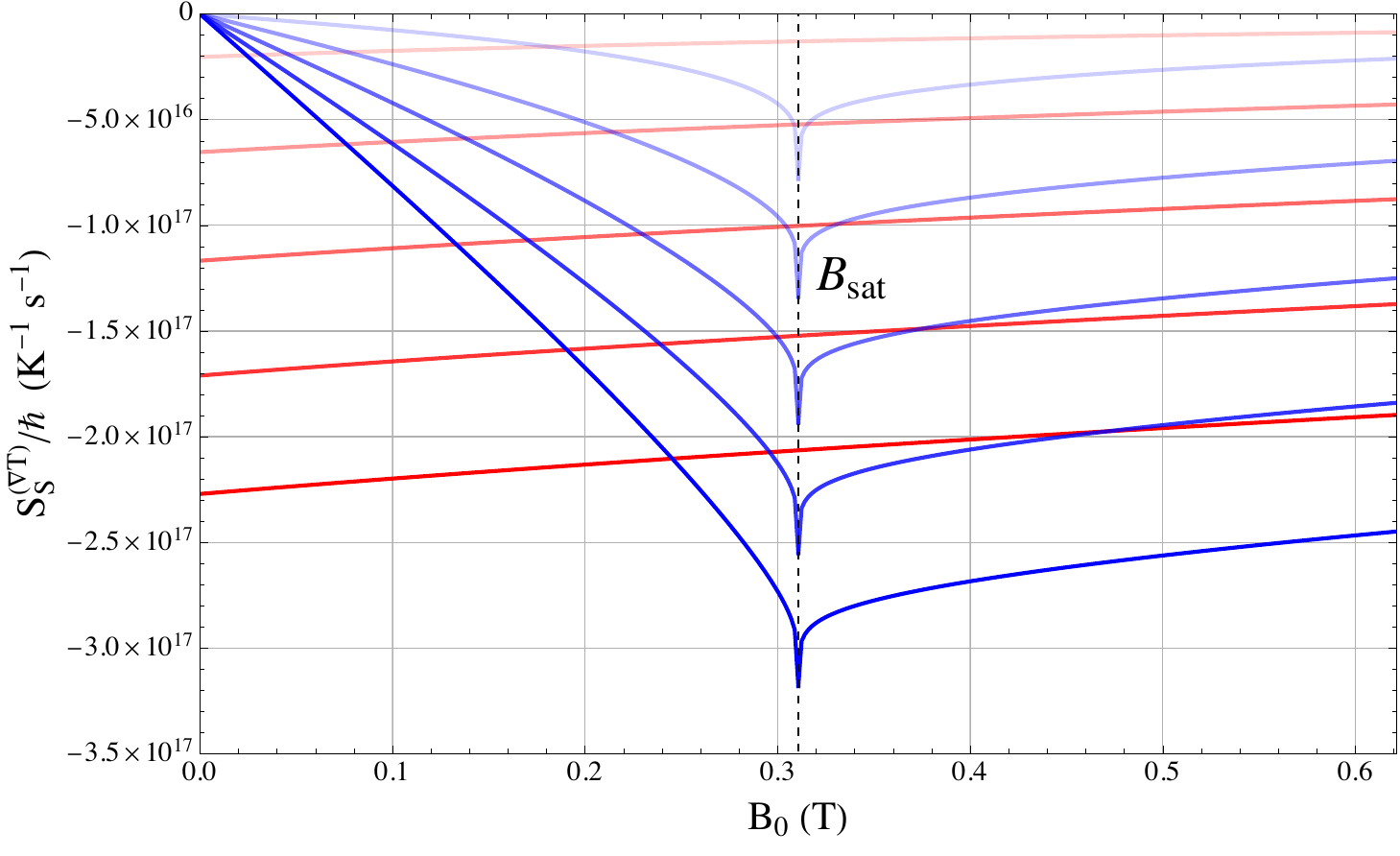}%
  %}
  \caption{\small Thermal SSC for various temperatures in a monolayer as a function of the external magnetic field \(B_0\). The red curves show \(S_S^{(\nabla T)}\) along the easy axis for $\vec{B}_0 \parallel \hat{b}$, the blue curves along the intermediate axis in the canted (\(B_0 < B_{\text{sat}}\)) and saturated (\(B_0 \geq B_{\text{sat}}\)) phases for $\vec{B}_0 \parallel \hat{a}$. The temperature ranges from \(T = 1\) K (most transparent curve) to \(T = 5\) K (opaque curve) in steps of \(1\) K.}
\end{figure}

Fig. \hyperref[fig12]{12} shows the thermal SSC in the monolayer with temperatures ranging from \(T = 1\) K (most transparent curve) to \(T = 5\) (opaque curve) K in steps of \(1\) K.

\begin{figure}[H] \label{fig13}
  %\subfloat{%
    \hspace{-3mm}\includegraphics[clip,width=1\columnwidth]{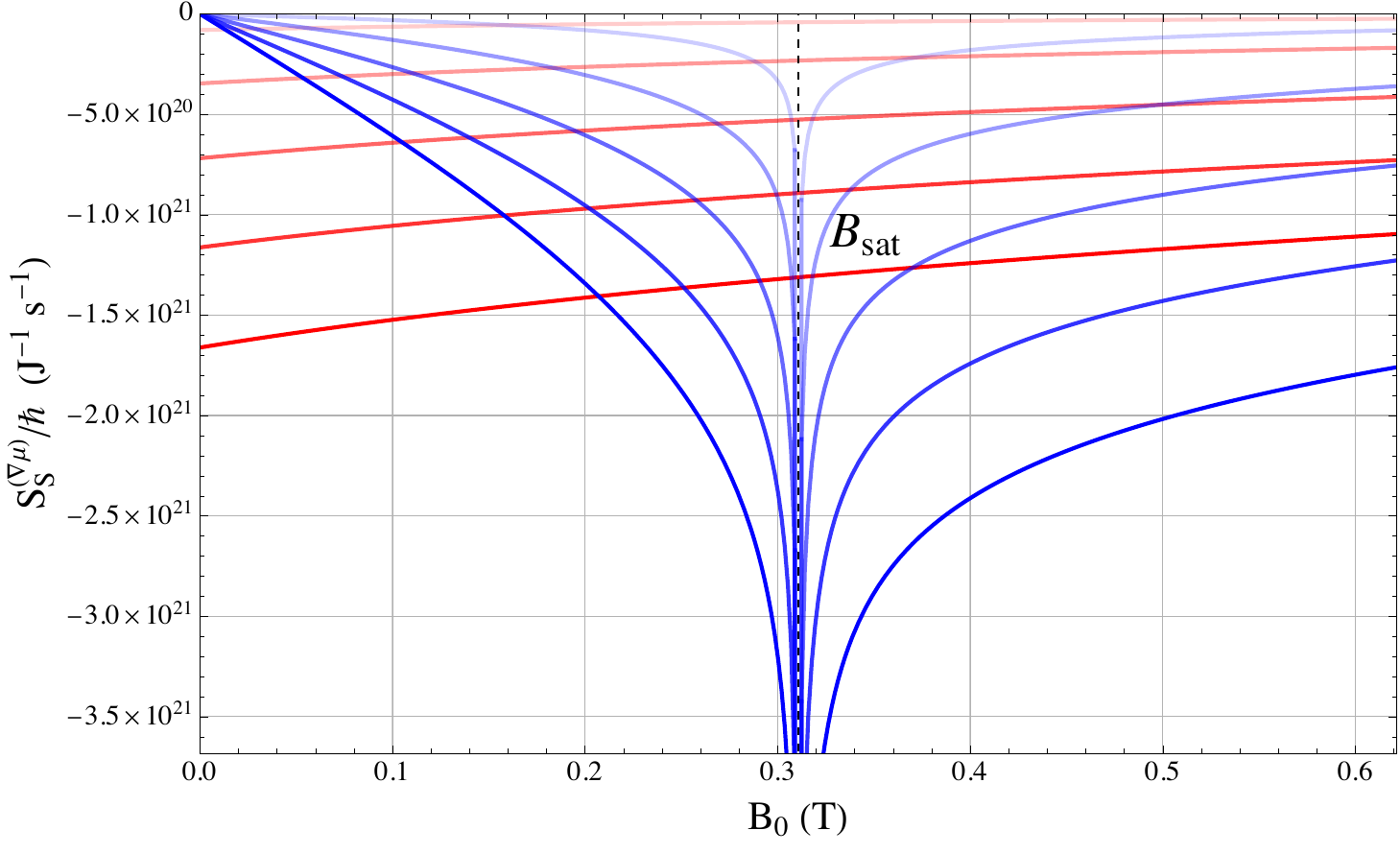}%
  %}
  \caption{\small Diffusive SSC for various temperatures in a monolayer as a function of the external magnetic field \(B_0\). The red curves show \(S_S^{(\nabla \mu)}\) along the easy axis for $\vec{B}_0 \parallel \hat{b}$, the blue curves along the intermediate axis in the canted (\(B_0 < B_{\text{sat}}\)) and saturated (\(B_0 \geq B_{\text{sat}}\)) phases for $\vec{B}_0 \parallel \hat{a}$. The temperature ranges from \(T = 1\) K (most transparent curve) to \(T = 5\) K (opaque curve) in steps of \(1\) K.}
\end{figure}

Fig. \hyperref[fig13]{13} shows the diffusive SSC in the monolayer with temperatures ranging from \(T = 1\) K (most transparent curve) to \(T = 5\) (opaque curve) K in steps of \(1\) K.

\subsection{Bilayer} \label{appD2}

\begin{figure} [h] \label{fig14}

%\subfloat[]{%
  \hspace{-3mm}\includegraphics[clip,width=1\columnwidth]{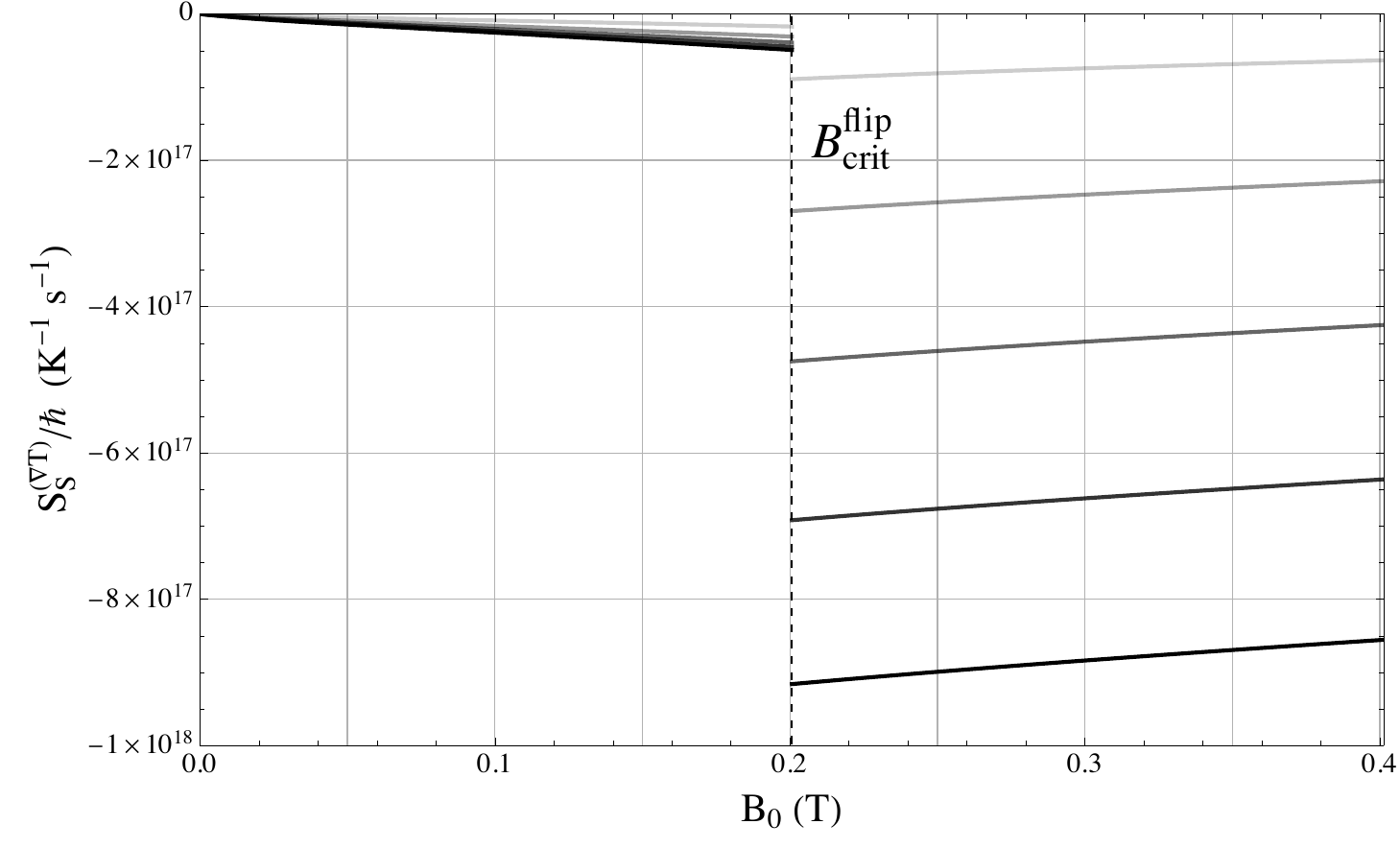}%}
\label{fig14a}

%\subfloat[]{%
  \hspace{-3mm}\includegraphics[clip,width=0.985\columnwidth]{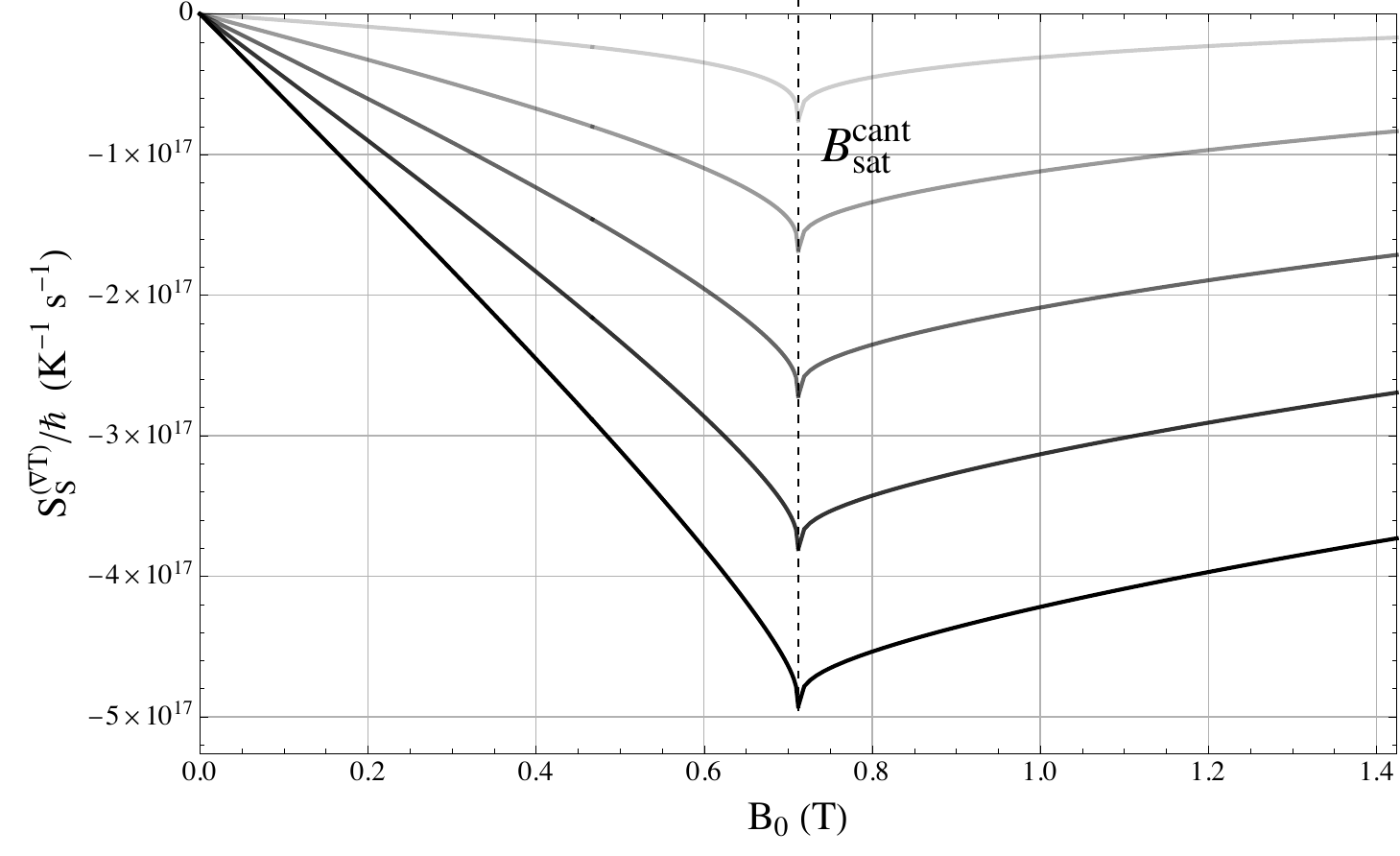}%}
  \label{fig14b}
\caption{ \small Thermal SSC of the sum of the two magnon modes for various temperatures in the bilayer as a function of the external magnetic field \(B_0\). The temperature ranges from \(T = 1\) K (most transparent curves) to \(T = 5\) K (opaque curves) in steps of \(1\) K. (a) \(S_S^{(\nabla T)}\) along the easy axis for $\vec{B}_0 \parallel \hat{b}$ in the AFM (\(B_0 < B^{\text{flip}}_{\text{crit}}\)) and FM (\(B_0 \geq B^{\text{flip}}_{\text{crit}}\)) phases. (b) \(S_S^{(\nabla T)}\) along the intermediate axis for $\vec{B}_0 \parallel \hat{a}$ in the canted (\(B_0 < B^{\text{cant}}_{\text{sat}}\)) and saturated (\(B_0 \geq B^{\text{cant}}_{\text{sat}}\)) phases.}
\end{figure}

Figs. \hyperref[fig14a]{14(a)} (AFM-FM) and \hyperref[fig14b]{14(b)} (canted-saturated) show the thermal SSC in the bilayer with temperatures ranging from \(T = 1\) K (most transparent curves) to \(T = 5\) (opaque curves) K in steps of \(1\) K.

\begin{figure} [h] \label{fig15}

%\subfloat[]{%
  \hspace{-3mm}\includegraphics[clip,width=1\columnwidth]{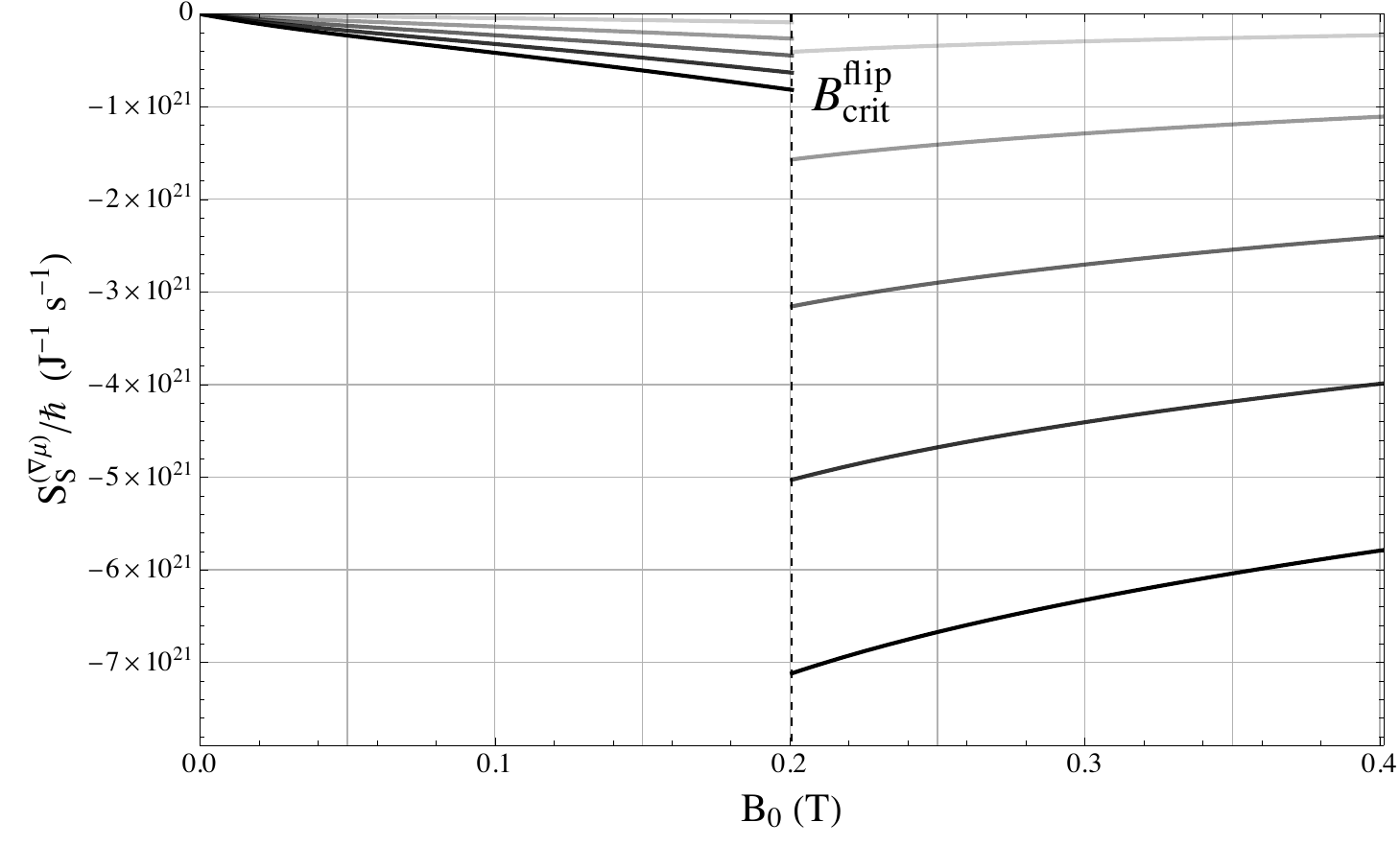}%}
\label{fig15a}

%\subfloat[]{%
  \hspace{-3mm}\includegraphics[clip,width=0.985\columnwidth]{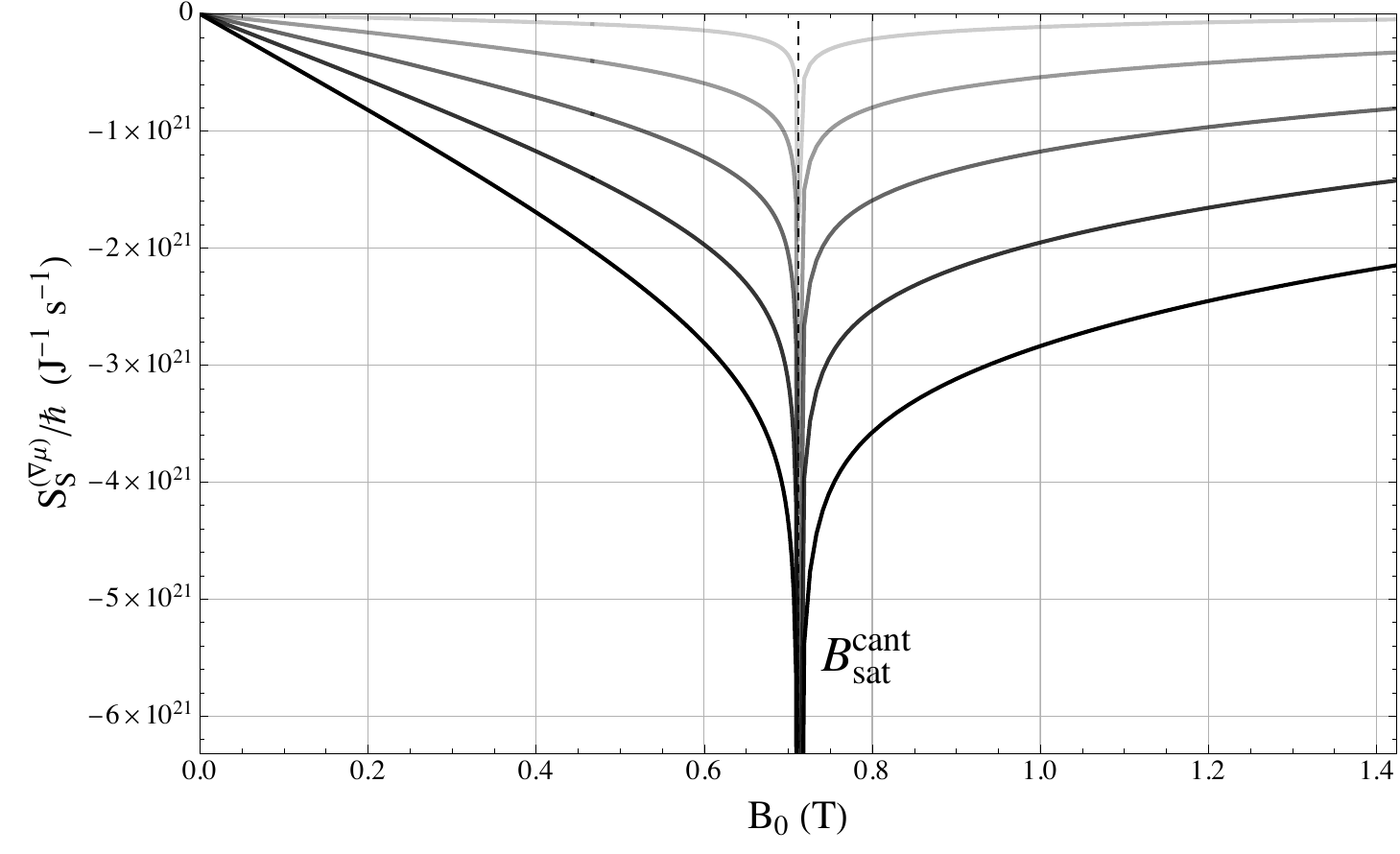}%}
  \label{fig15b}
\caption{ \small Diffusive SSC of the sum of the two magnon modes for various temperatures in the bilayer as a function of the external magnetic field \(B_0\). The temperature ranges from \(T = 1\) K (most transparent curves) to \(T = 5\) K (opaque curves) in steps of \(1\) K. (a) \(S_S^{(\nabla \mu)}\) along the easy axis for $\vec{B}_0 \parallel \hat{b}$ in the AFM (\(B_0 < B^{\text{flip}}_{\text{crit}}\)) and FM (\(B_0 \geq B^{\text{flip}}_{\text{crit}}\)) phases. (b) \(S_S^{(\nabla \mu)}\) along the intermediate axis for $\vec{B}_0 \parallel \hat{a}$ in the canted (\(B_0 < B^{\text{cant}}_{\text{sat}}\)) and saturated (\(B_0 \geq B^{\text{cant}}_{\text{sat}}\)) phases.}
\end{figure}

Figs. \hyperref[fig15a]{15(a)} (AFM-FM) and \hyperref[fig15b]{15(b)} (canted-saturated) show the diffusive SSC in the bilayer with temperatures ranging from \(T = 1\) K (most transparent curves) to \(T = 5\) (opaque curves) K in steps of \(1\) K.

\section{Magnon thermal conductivity} \label{appE}

Here we show the magnon thermal conductivity in the mono- and bilayer. The heat flux flowing in the \(\eta\)-direction is

\begin{equation}\label{E1}
\begin{aligned}
        J^{(\eta,\varepsilon)}_{S} & = -\frac{\tau}{(2 \pi)^2}\int \hbar \omega_{\vec{k}}  v^2_{k,\eta}\frac{\partial n^0}{\partial T}d^2k \frac{\partial T}{\partial \eta} \\
        & \equiv - \kappa(B_0) \frac{\partial T}{\partial \eta}, \\
\end{aligned}
\end{equation}
where we introduced the magnon thermal conductivity \(\kappa(B_0)\) .

\subsection{Monolayer} \label{appE1}

\begin{figure}[h] \label{fig16}
  %\subfloat{%
    \hspace{-3mm}\includegraphics[clip,width=1\columnwidth]{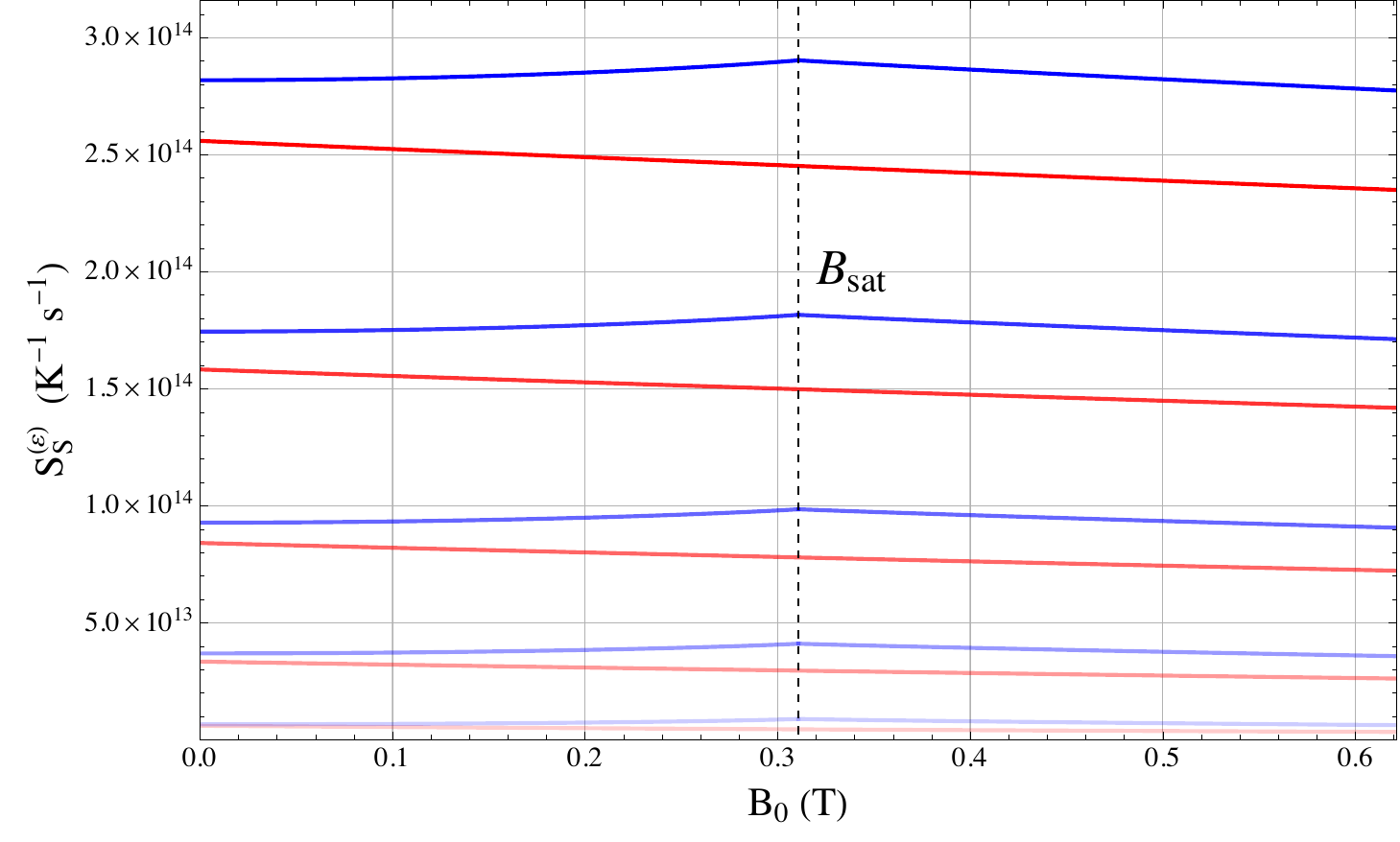}%
  %}
  \caption{\small Magnon thermal conductivity for various temperatures in a monolayer as a function of the external magnetic field \(B_0\). The temperature ranges from \(T = 1\) K (most transparent curves) to \(T = 5\) K (opaque curves) in steps of \(1\) K. The red curves show \(\kappa\) for $\vec{B}_0 \parallel \hat{b}$, the blue curves in the canted (\(B_0 < B_{\text{sat}}\)) and saturated (\(B_0 \geq B_{\text{sat}}\)) phases for $\vec{B}_0 \parallel \hat{a}$.}
\end{figure}

Fig. \hyperref[fig16]{16} shows the magnon thermal conductivity in the monolayer with temperatures ranging from \(T = 1\) K (most transparent curve) to \(T = 5\) (opaque curve) K in steps of \(1\) K.

\subsection{Bilayer} \label{appE2}

\begin{figure} [H] \label{fig17}

%\subfloat[]{%
  \hspace{-3mm}\includegraphics[clip,width=1\columnwidth]{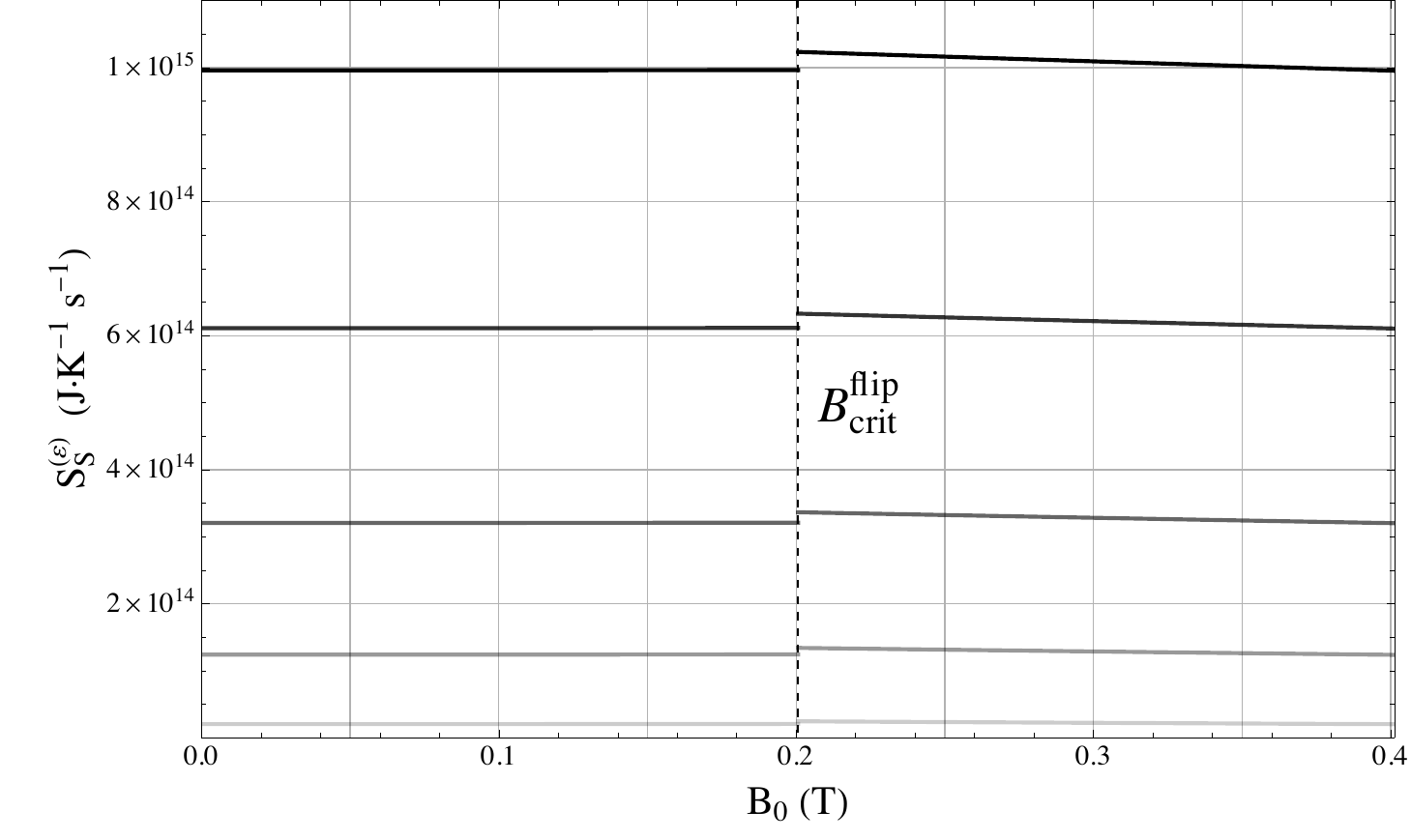}%}
\label{fig17a}

%\subfloat[]{%
  \hspace{-3mm}\includegraphics[clip,width=1\columnwidth]{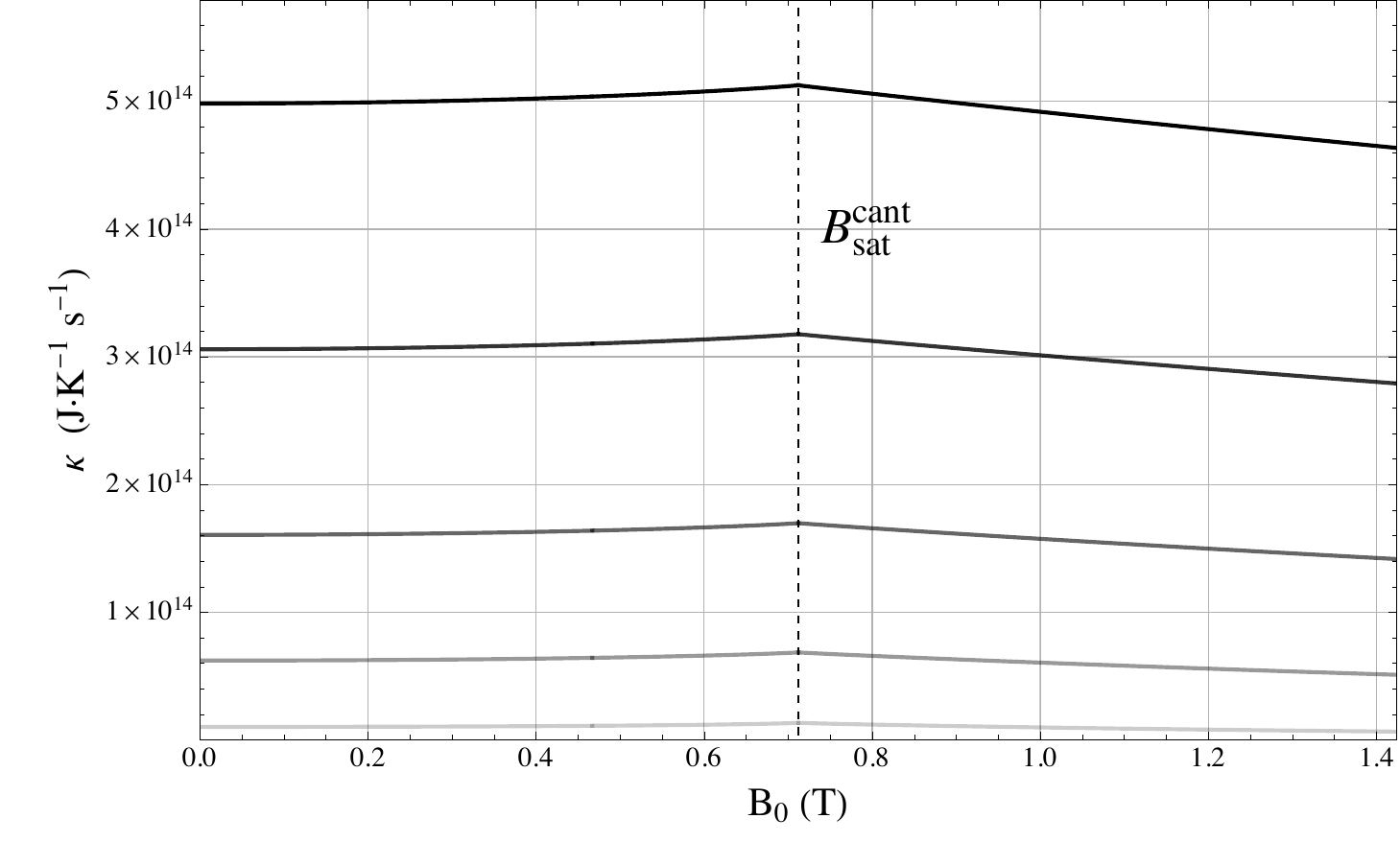}%}
  \label{fig17b}
\caption{ \small Magnon thermal conductivity of the sum of the two magnon modes for various temperatures in the bilayer as a function of the external magnetic field \(B_0\). The temperature ranges from \(T = 1\) K (most transparent curves) to \(T = 5\) K (opaque curves) in steps of \(1\) K. (a) \(\kappa\) for $\vec{B}_0 \parallel \hat{b}$ in the AFM (\(B_0 < B^{\text{flip}}_{\text{crit}}\)) and FM (\(B_0 \geq B^{\text{flip}}_{\text{crit}}\)) phases. (b) \(\kappa\) for $\vec{B}_0 \parallel \hat{a}$ in the canted (\(B_0 < B^{\text{cant}}_{\text{sat}}\)) and saturated (\(B_0 \geq B^{\text{cant}}_{\text{sat}}\)) phases.}
\end{figure}

Figs. \hyperref[fig17a]{17(a)} (AFM-FM) and \hyperref[fig17b]{17(b)} (canted-saturated) show the magnon thermal conductivity in the bilayer with temperatures ranging from \(T = 1\) K (most transparent curves) to \(T = 5\) (opaque curves) K in steps of \(1\) K.

\bibliography{apssamp}% Produces the bibliography via BibTeX.

\end{document}